\documentclass[12pt,a4]{paper}
\usepackage{axodraw,a4}
\usepackage{minitoc,cite,fancyhdr}
\usepackage{bbm} 
\usepackage{euscript,amssymb,latexsym} 
\usepackage{epsf,epsfig,graphics,subfigure,graphicx}
\usepackage{amsmath}
\usepackage{multirow}

\usepackage{eepic}
             
\makeatletter
\def\fmslash{\@ifnextchar[{\fmsl@sh}{\fmsl@sh[0mu]}}
\def\fmsl@sh[#1]#2{%
  \mathchoice
    {\@fmsl@sh\displaystyle{#1}{#2}}%
    {\@fmsl@sh\textstyle{#1}{#2}}%
    {\@fmsl@sh\scriptstyle{#1}{#2}}%
    {\@fmsl@sh\scriptscriptstyle{#1}{#2}}}
\def\@fmsl@sh#1#2#3{\m@th\ooalign{$\hfil#1\mkern#2/\hfil$\crcr$#1#3$}}
\makeatother
\global\arraycolsep=2pt 
\def\be{\begin{equation}}
\def\ee{\end{equation}}
\newcommand{\bea}{\begin{eqnarray}}
\newcommand{\eea}{\end{eqnarray}}

\begin{document}
\begin{flushright}
IPPP/11/50\\
DCPT/11/100\\
\end{flushright}
{\Large \bf Simulation of Sextet Diquark Production}\\[0.7cm]
\hspace{-16pt}{\bf\normalsize{Peter Richardson, David Winn}}
\\ 
{\it\normalsize{Institute of Particle Physics Phenomenology, Department of Physics}}
\\ 
{\it\normalsize{University of Durham, DH1 3LE, UK;}}
\\
{\it\normalsize{Email:} }
{\sf\normalsize{peter.richardson@durham.ac.uk, d.e.winn@durham.ac.uk}}
\begin{abstract}
We present a method for simulating the production and decay of
particles in the sextet representation of $SU(3)_C$ including
the simulation of QCD radiation. Results from the 
Monte Carlo simulation of sextet
diquark production at the LHC including both resonant and pair
production are presented.
We include limits on resonant diquark production from
recent ATLAS results and perform the first simulation studies of the less model
dependent pair production mechanism.
\end{abstract}
\thispagestyle{empty}
\thispagestyle{plain}
\setcounter{page}{1}
\noindent
---------------------------------------------------------------------------------------------------------
\section{Introduction}

Many models of Beyond the Standard Model~(BSM) physics
require the inclusion of diquarks. For
example, diquarks appear in a number of Grand Unified Theories (GUTs)
and have even been postulated as a form of dynamical symmetry
breaking, giving rise to the masses of particles,
\cite{Marciano:1980zf,Fukazawa:1990fb}. The colour sextet diquark is, in 
group theory language, a rank 2 symmetric tensor formed from
the direct product of two fundamental representations
$\mathbf{3} \otimes \mathbf{3} = \mathbf{6} \oplus \mathbf{\bar{3}}$.
As such it is the lowest colour representation which has not been
observed and therefore investigation of sextet diquark production
at the CERN Large Hadron Collider (LHC) is interesting in its own right.

The LHC experiments have started data taking 
at the high energy frontier ($\sqrt{s}=\mathrm{7 \, TeV}$), allowing probes of
energy scales not previously seen. 
At the LHC because the fundamental collisions are between the
quarks and gluons inside the colliding protons 
the strong force is the dominant interaction 
allowing Quantum Chromodynamics~(QCD) to be studied at these new high
energies.
As a diquark is produced via strong interactions, and with the
potential of a relatively low mass, diquarks may be seen in the early
stages of LHC data taking.
The LHC also favours the formation of
diquarks from the valence quarks as it is
a proton-proton collider as opposed to a
proton-antiproton collider, such as the Tevatron.

Due to their exotic colour structure and
$SU(3)_C$ quantum numbers, diquarks will give rise to jets in
the detector. The expected signals will be either a resonance in the
invariant dijet mass distribution or the production of two equal mass 
dijet systems in four jet events in the case of pair production.
In order to study the
experimental signatures of diquark production, a Monte Carlo simulation is
required that includes the production of sextet particles, their perturbative
decays and the full Monte Carlo machinery of showering (including the
exotic colour structure) and hadronization.

Although significant efforts have been made to study the resonant
production of diquarks \cite{Han:2009ya,
  Zhang:2010kr,Chen:2008hh,Arik:2001bc,Mohapatra:2007af,Berger:2010fy,Cakir:2005iw}
and also pair production\cite{Tanaka:1991nr,Chen:2008hh,Patel:2011eh},
a full study of experimental signatures including Monte Carlo
simulations has not been performed. In this paper, we discuss: the
implementation of the diquark model in general purpose Monte Carlo
event generators; place constraints on the coupling as a function of
mass based on the latest ATLAS
results~\cite{Aad:2011aj,ATLAS-CONF-2011-081}; present some results of
invariant mass distributions both for resonant and pair production.

\section{Simulation}

Monte Carlo simulations describe high energy collisions using:
\begin{enumerate}
\item a hard perturbative, 
either leading- or next-to-leading-order, matrix element
to simulate the fundamental hard collision process;
\item the parton shower algorithm which evolves from the scale
of the hard process to a cut-off scale, $\mathcal{O}(1\,{\rm GeV})$,
via the successive radiation of soft and collinear quarks and gluons;
\item the generation of multiple perturbative scattering
      processes to simulate the underlying event;
\item the perturbative decay of  any fundamental particles, with
      lifetimes shorter than the timescale for hadron formation, followed
      again by the simulation of QCD radiation from the coloured decay products
      using the parton shower formalism;
\item a multiple partonic collision model is used to simulate the underlying event; 
\item a hadronization model which describes the formation
      of hadrons at the cut-off scale from the quarks and gluons produced
      during the parton shower;
\item the decays of the unstable hadrons produced by hadronization.
\end{enumerate}
The various calculations, approximations and models used in these
simulations are reviewed in Ref.~\cite{Buckley:2011ms}.  Simulating
most models of BSM physics only requires the implementation of the
various hard production and decay processes with the simulation of
perturbative QCD radiation and hadronization proceeding in the same
way as in the Standard Model~(SM), provided the new particles decay before
forming hadrons. However, in models including the production and decay of
sextet particles or R-parity violating SUSY\footnote{The simulation of
  R-parity violating SUSY models was considered in detail in
  Ref.\,\cite{Dreiner:1999qz}.}  the new colour structures require
changes to both the simulation of QCD radiation and the subsequent
hadronization.

As with all models the simulation starts with the calculation of the
hard production and decay processes using the most general Lagrangian
for the coupling of the sextet particles to the
quarks~\cite{Cakir:2005iw,Zhang:2010kr,Atag:1998xq,Arik:2001bc,Ma:1998pi}
\be
\begin{aligned}
\label{eq:lag}
\mathcal{L}= & 
\left(g_{1L}\overline{q_{L}^{c}}i\tau_{2}q_{L}+g_{1R}\overline{u_{R}^{c}}d_{R}\right)\Phi_{1,1/3} 
\:+\: g_{1R}^{\prime}\overline{d_{R}^{c}}d_{R}\Phi_{1,-2/3}
\:+\: g_{1R}^{\prime\prime}\overline{u_{R}^{c}}u_{R}\Phi_{1,4/3} \:+ \\
& g_{3L}\overline{q_{L}^{c}}i\tau_{2}\tau q_{L}\cdot\Phi_{3,1/3}
\:+\: g_{2}\overline{q_{L}^{c}}\gamma_{\mu}d_{R}V_{2,-1/6}^{\mu}
\:+\: g_{2}^{\prime}\overline{q_{L}^{c}}\gamma_{\mu}u_{R}V_{2,5/6}^{\mu}
\:+\: h.c. \, ,
\end{aligned}
\ee
where $q_L$ is the left-handed quark doublet, $u_{R}$ and $d_{R}$ are the
right-handed quark singlet fields, and $q^{c}\equiv C\bar{q}^{T}$ is
the charge conjugate quark field. The colour and generation indices
are omitted to give a more compact notation and the subscripts on the
scalar, $\Phi$, and vector, $V^{\mu}$, fields denote the SM electroweak
gauge quantum numbers: ($SU(2)_L$, $U(1)_Y$). The 
Lagrangian is
assumed to be flavour diagonal to avoid any flavour
changing currents arising from the new interactions. 

The kinetic and QCD terms in the Lagrangian are
\begin{subequations}
\be
\mathcal{L}^{\rm scalar}_{\rm QCD} = D^\mu\Phi D_\mu\Phi -m^2\Phi\Phi^\dagger,
\ee
for scalar diquarks, where $\Phi$ is the scalar diquark field and
\be
\mathcal{L}^{\rm vector}_{\rm QCD} = -\frac14\left(D^\mu V^\nu-D^\nu V^\mu\right)\left(D_\mu V_\nu-D_\nu V_\mu\right) -m^2V^\mu V_\mu,
\ee
\end{subequations}
for vector diquarks, where $V^\mu$ is the vector diquark field. The covariant
derivative $D^\mu$ has the standard form for Quantum Chromodynamics.

The simulation of perturbative QCD radiation, relies on the large number of
colours, $N_C$, limit for both the treatment of perturbative QCD radiation
and the subsequent hadronization. In this approach particles in
 the fundamental representation of $SU(N_C)$ carry a colour, those
in the antifundamental representation an anticolour and
those in the adjoint representation both a colour and an anticolour.
This allows us to consider the flow of colour, via colour lines,
 in hard interactions which is determined by the colour structure of
the hard perturbative matrix elements.

Using this colour flow both:
\begin{itemize}
\item soft gluon radiation, where in a given hard process there 
      is a maximum angle for the radiation of gluon related to the colour
      flow of the process;
\item and hadronization, where the first step of both the
      string~\cite{Andersson:1983ia,Andersson:1998tv} and 
      cluster~\cite{Webber:1983if} models is the formation of colour singlet systems;
\end{itemize}
can be simulated.

This is complicated in models involving sextet particles where in the
large-$N_C$ limit the sextet particles possess two fundamental
colours, appropriately symmetrized.  This cannot be handled by
conventional Monte Carlo simulations which require all the colours of
the particles to have fundamental colours and/or anticolours.  In
order to simulate these particles we choose to represent (anti)sextet
particles as having two (anti)colours. 

Consider the production and
subsequent decay of a scalar sextet particle. In order to simulate QCD
radiation from the intermediate sextet resonance we have to simulate
the production and decay separately. The matrix element for the
process is
\be
\mathcal{M}=\mathcal{M}_{i{\rm prod}} \frac{i\delta^i_j}{p^2-m^2} \mathcal{M}_{\rm decay}^j
\ee
where $i,j$ are colour indices of the sextet particle, $\mathcal{M}_{i{\rm prod}}$ is 
the matrix element for the production of a scalar sextet particle with colour $i$,
four-momentum $p$ and mass $m$,
and $\mathcal{M}_{\rm decay}^j$ is the matrix element for the decay of
a scalar sextet particle with colour $j$.

This can be rewritten using $\delta^i_j=K^i_{ab}\bar{K}_j^{ba}$ where
$K$ and $\bar{K}$ are the Clebsch-Gordan coefficients in the sextet and antisextet
representations, respectively. Hence
\be
\mathcal{M}
= \mathcal{M}_{i{\rm prod}} \frac{iK^i_{ab}\bar{K}_j^{ba}}{p^2-m^2} \mathcal{M}_{\rm decay}^j
= \mathcal{M'}_{ab{\rm prod}} \frac{i}{p^2-m^2} \mathcal{M'}_{\rm decay}^{ba}.
\ee
In order to consider the intermediate sextet particle as having two fundamental
colours we have absorbed the Clebsch-Gordan into the redefined production,
$\mathcal{M'}_{ab{\rm prod}}$, and decay matrix elements, $\mathcal{M'}_{\rm decay}^{ba}$.

\begin{figure}[t]
\begin{center}
  \includegraphics[width=0.6\textwidth]{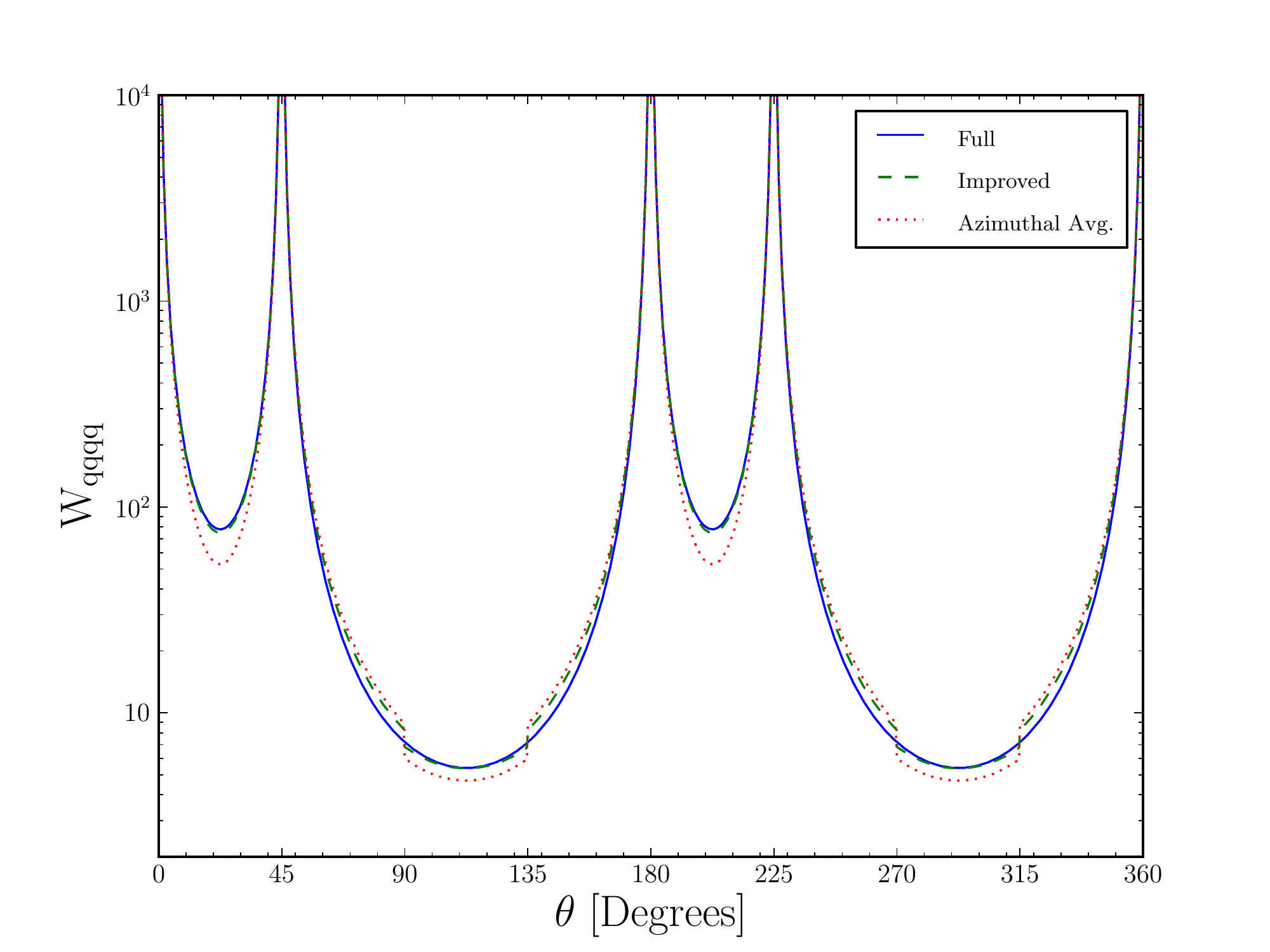}
\end{center}
  \caption{The radiation pattern of associated with gluon emission
    from the incoming and outgoing quarks during resonant production, 
    where $\theta$ is the polar angle of the gluon with respect to the
    $z$-axis.}
  \label{fig:radPattern}
\end{figure}

From this, the colour partners of the decay products can be determined and the 
usual angular ordering procedure applied ~\cite{Marchesini:1987cf, Buckley:2011ms}. 
The radiation pattern of gluons from the quarks for the resonant
production of diquarks,\linebreak \mbox{$q(p_1)q(p_2)\to\{\Phi,V\}\to q(p_3)q(p_4)$}, is
\begin{align}
J^2 = C_F \left[\rule{0mm}{6mm}\right.&(W_{13}+W_{14}+W_{23}+W_{24}) \label{eq:radPattern}\\
    &\left. + \frac{2}{N_C+1}\left[\frac{1}{2}(W_{13}+W_{14}+W_{23}+W_{24}) - W_{12} - W_{34}\right]\right].
 \nonumber 
\end{align}
The dipole radiation function for two massless particles $i$ and $j$ radiating
a gluon is
\begin{equation}
W_{ij} = \omega^2 \frac{p_i\cdot p_j}{(p_i\cdot k)(p_j\cdot k)} = W^i_{ij}+W^j_{ij},
\end{equation}
where $p_{i,j}$ are the 4-momenta of the radiating particles and, $k$ and
$\omega$ are the 4-momentum of energy of the gluon, respectively.
In terms of the angle between the gluon and particle $i$, $\theta_i$,
and the angle between the particles $i$ and $j$, $\theta_{ij}$,
\begin{equation}
W^i_{ij} = \frac{1}{2(1-\cos \theta_{i})}\left( 1 + \frac{\cos \theta_i - \cos \theta_{ij}}{1-\cos \theta_{i}}\right).
\end{equation}
The last term in Eq.~\ref{eq:radPattern} can be
neglected, as usual, due to both the $\tfrac{1}{N_C}$ suppression, compared to
the leading term, and the dynamical suppression in the massless limit because there is no collinear singularity
in this term. The radiation pattern
is shown in Figure~\ref{fig:radPattern}, where the outgoing quarks
were held at $45^{\circ}$ and $225^{\circ}$ with respect to the
incoming beam direction. The full radiation pattern, the result
after neglecting the subleading terms and azimuthally averaging, and
the improved angular ordered result, where the full result is used instead of 
the azimuthal average inside the angular-ordered region, are shown.
Improved angular ordering, as
implemented in \textsf{Herwig++} performs well in the collinear limit.

The details of the colour decomposition for
both the resonant and pair production of sextet particles
and the radiation of gluons from sextet particles are
presented in Appendix~\ref{app:shower}. The same approach was recently used in
{\sc{MadGraph}}~5~\cite{Alwall:2011uj} including the ability to
automatically decompose the colours for higher representations of
the gauge group. They also study the impact of matching the hardest
perturbative emission in resonance production
which we have not considered but will not effect on our results.

\section{Phenomenology}

 In order to study the phenomenology simulations were performed for
 the  scalar $\Phi_{1,4/3}$ and the 
 vector $V^{\mu +}_{2,5/6}$ diquarks
 which were chosen as they can be produced as $s$-channel resonances
 from the partonic collision of the valence up quarks.

\begin{figure}[t!!]
\begin{center}
  \subfigure[Scalar]{ 
    \includegraphics[width=0.45\textwidth]{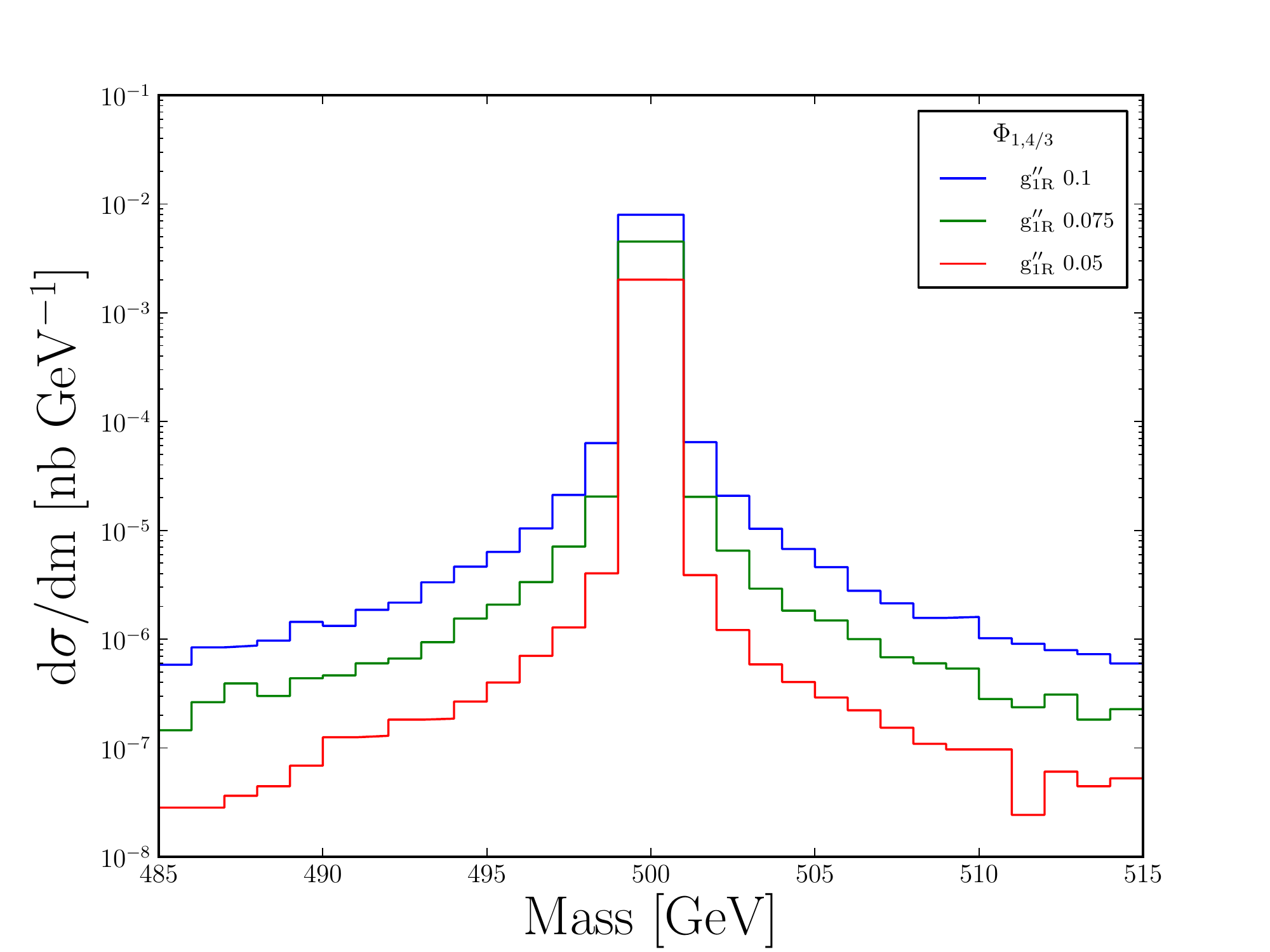}
    \label{subfig:scalarBreit}
  }
  \subfigure[Vector]{
    \includegraphics[width=0.45\textwidth]{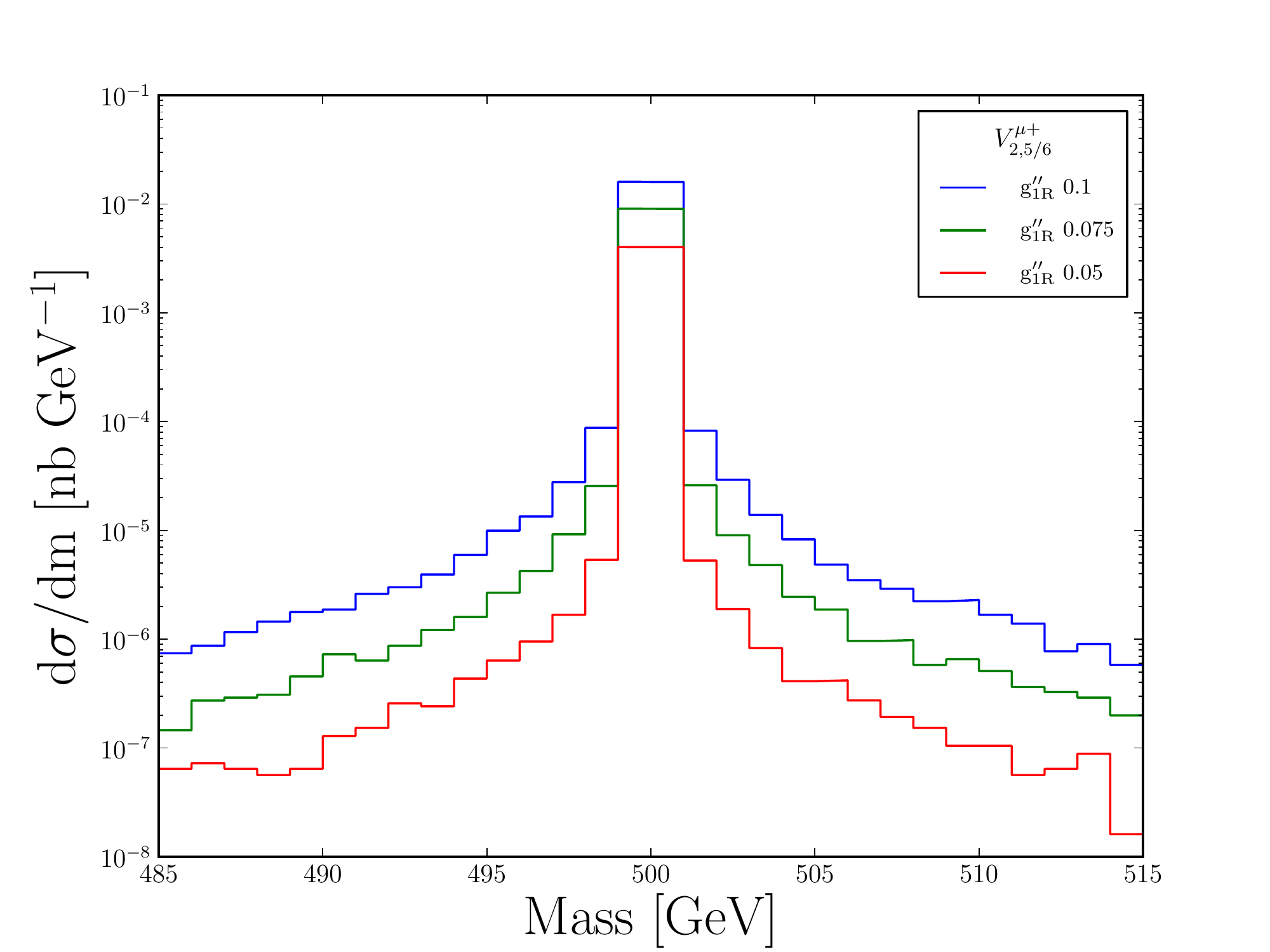}
    \label{subfig:vectorBreit}
  }
\end{center}
\caption{The Breit-Wigner shapes for $\sqrt{s} = \mathrm{14\,TeV}$ with couplings
  of 0.1, 0.075 and 0.05.}
\label{fig:breit}
\begin{center}
  \includegraphics[width=0.6\textwidth]{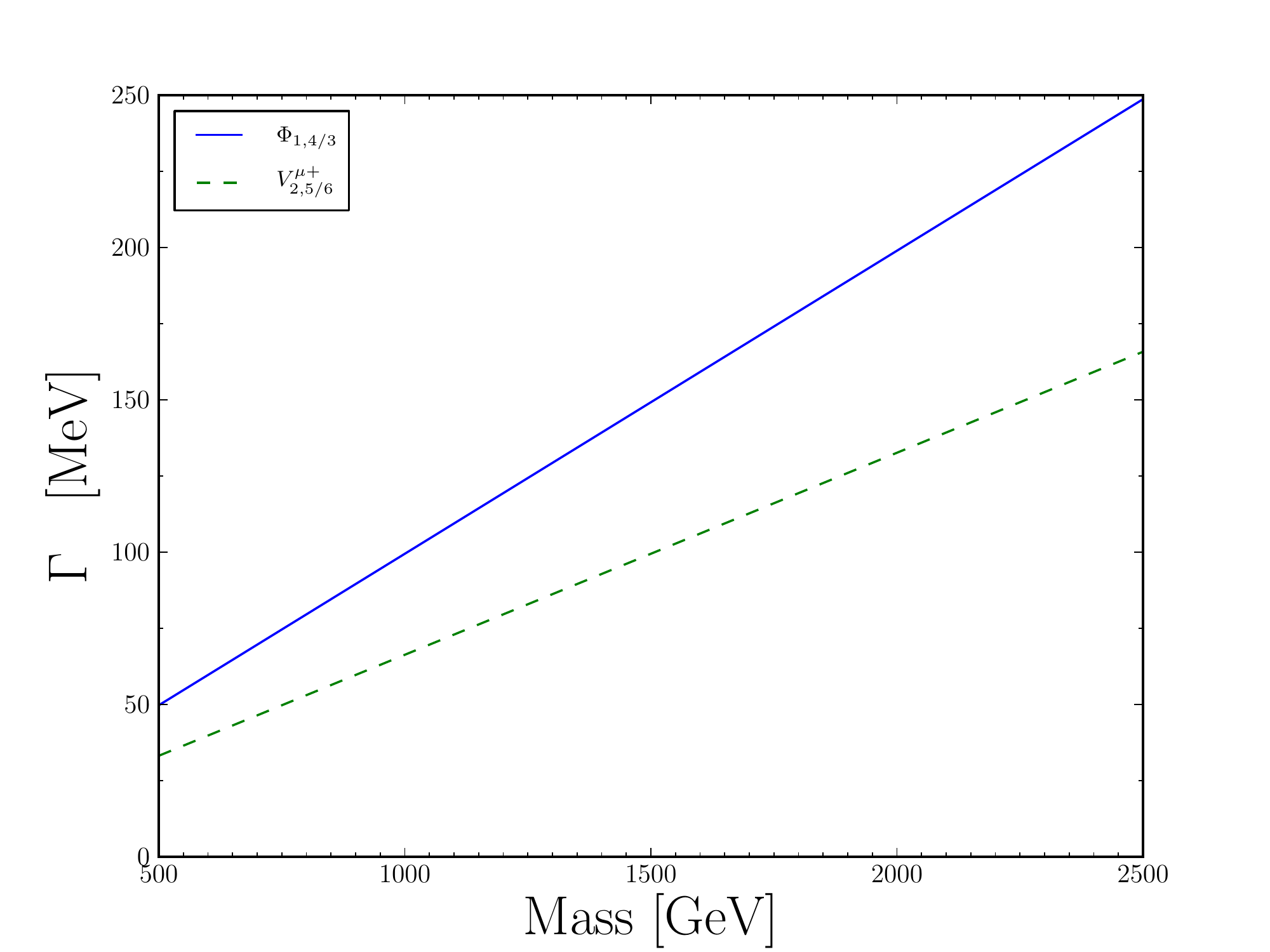}
\end{center}
\caption{The diquark width as a function of the diquark mass for a coupling
         as quoted in the text for scalar and vector diquarks. The diquark coupling
         to quarks has been taken to be $0.1$.}
\label{fig:widthVsMass}
\end{figure}

  In all our analyses, jets were clustered using the
  anti-$k_T$ algorithm \cite{Cacciari:2008gp}, as implemented in
  the \textsf{FastJet} package~\cite{Cacciari:2005hq}, using a radius parameter
  of $R = 0.6$. This choice is typical for the ATLAS experiment 
  at the LHC~\cite{Aad:2011aj}. The LO$^{**}$ PDFs of Ref.\,~\cite{Sherstnev:2007nd},
  which are the default choice in \textsf{Herwig++}, were used.

 There are phenomenological constraints on the diquark couplings from
 \mbox{$D^0-\bar{D^0}$} mixing and non-strange pion decays~\cite{Mohapatra:2007af}.
 For the up-type quarks there are constraints require that
\begin{equation}
g^{uu}_R \lesssim 0.1 \quad \mathrm{and} \quad g^{cc}_R \sim 0.
\end{equation}
The $g_L$ couplings have to be constrained due
to minimal flavour violation as the left-handed CKM matrix is well
known \cite{Han:2009ya}.

 It was therefore decided to take the
couplings 
\begin{equation}
g^{11}_{R/L} = 0.1 \quad \mathrm{and} \quad g^{22}_{R/L} = g^{33}_{R/L} = 0,
\end{equation}
where the numbered indices refer to the generation.

The value of the coupling will affect any studies involving jets as the
width of the particle varies as a function of the couplings. As seen in
Figure~\ref{fig:breit}, a larger coupling produces a larger width,
contributing to the smearing of a peak in the invariant mass
spectrum of the jets. If the coupling is less than the value chosen
above, any peak maybe enhanced compared to what is presented in the
following sections. The width as a function of the diquark mass is shown in
Figure~\ref{fig:widthVsMass}.

\subsection{Resonance Production}

\begin{figure}[t]
  \begin{center}
    \includegraphics[width=0.45\textwidth]{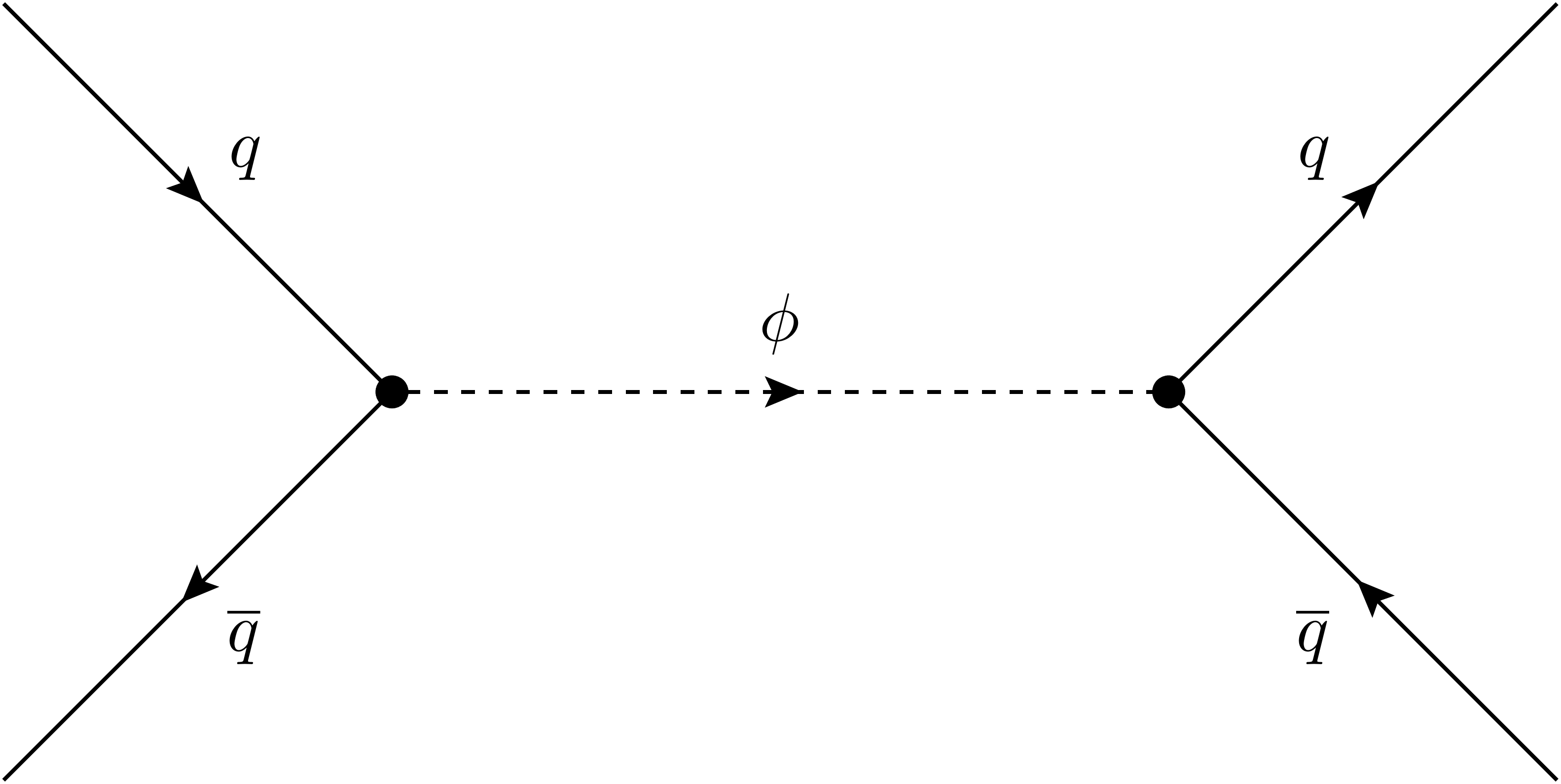}
    \caption{Resonant production and decay of a diquark}
    \label{fig:resProduction}
  \end{center}
\end{figure}

If the diquark has an appropriate mass and coupling it may be
resonantly produced at the LHC. The resonance production and
subsequent decay of a general diquark (scalar or vector) is shown in
Figure~\ref{fig:resProduction}, where the decay of the diquark will
depend on its mass and unknown couplings to the quarks, $g$.

\begin{figure}[t]
\begin{center}
\includegraphics[width=0.6\textwidth]{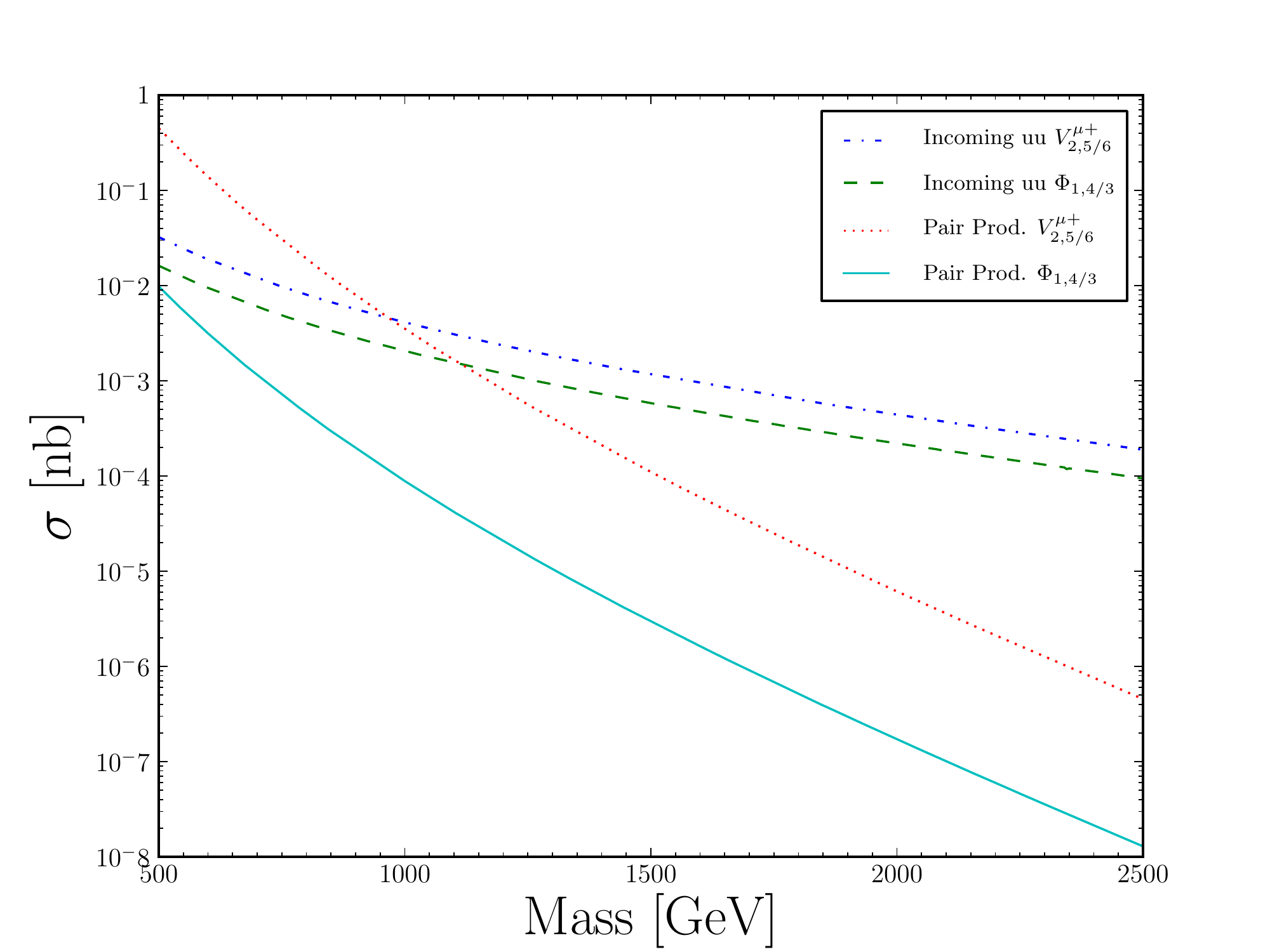}
\end{center}
\caption{Cross section for the production of
         vector and scalar diquarks as a function of the diquark mass for
         both resonant production, from incoming $uu$ states, and diquark 
         pair production at $\sqrt s =14$\,TeV. The diquark coupling to quarks
         has been taken to be $0.1$.}
\label{fig:crossSection}
\end{figure}

Figure~\ref{fig:crossSection} shows the cross section for the
production of scalar and vector diquarks from incoming uu quarks
(resonant production) and for incoming gluons (pair production).
The resonant production cross section depends quadratically on the
unknown diquark coupling to quarks,
which as been assumed to be $0.1$ in this
plot, whereas the pair production cross section is independent of this
coupling.

The diquark will decay into two quarks giving rise to at least two jets.
The search for the production of a diquark via resonant production
should therefore be in the dijet invariant mass spectrum,
where a smeared peak is
expected around the diquark mass. The primary background to this
search channel is QCD $2\rightarrow2$ scattering processes.

\begin{figure}[t!!]
\begin{center}
\subfigure[Scalar]{
  \includegraphics[width=0.47\textwidth]{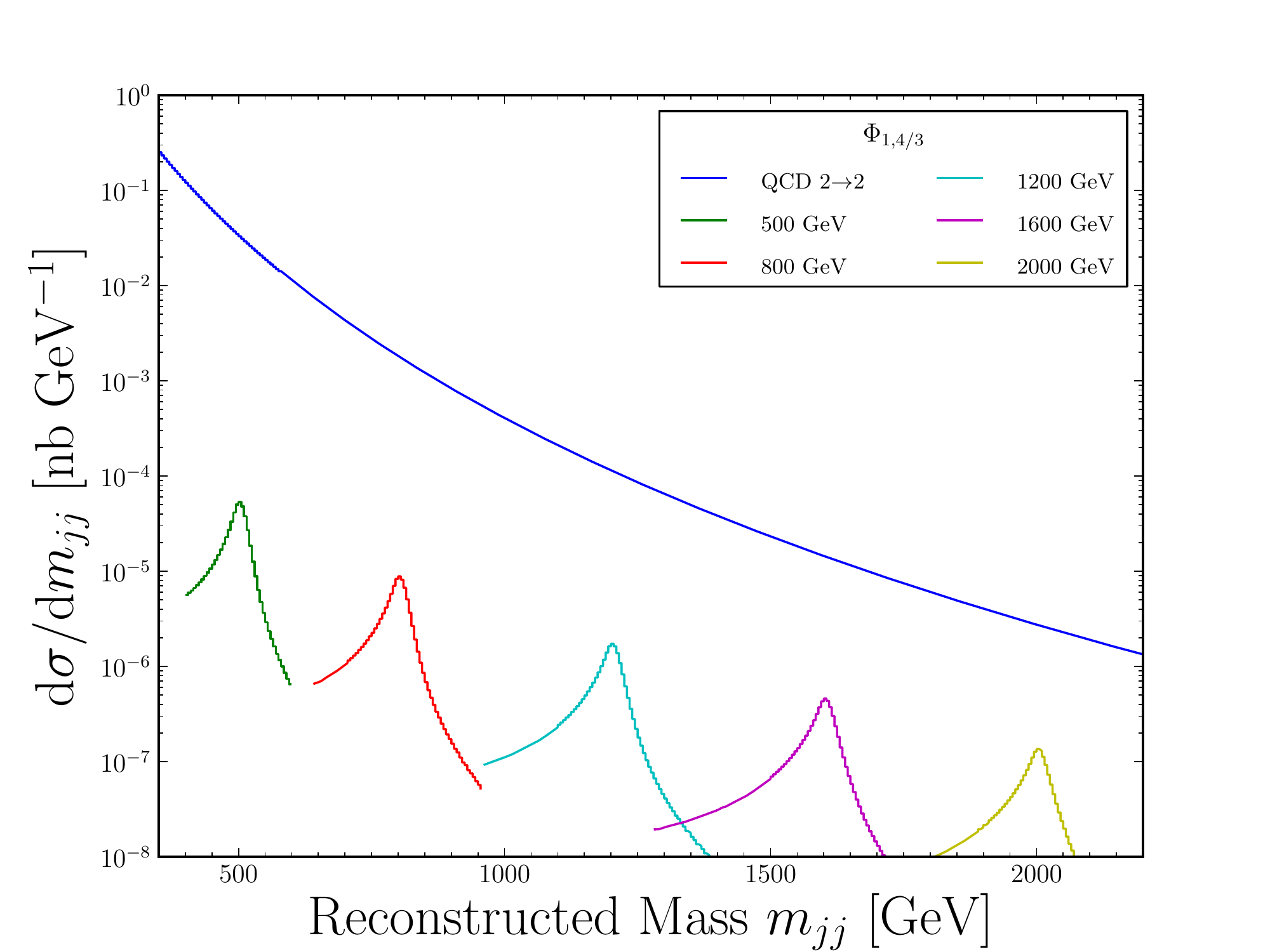}
  \label{subfig:diJetScalarMassA}
}
\subfigure[Vector]{ 
 \includegraphics[width=0.47\textwidth]{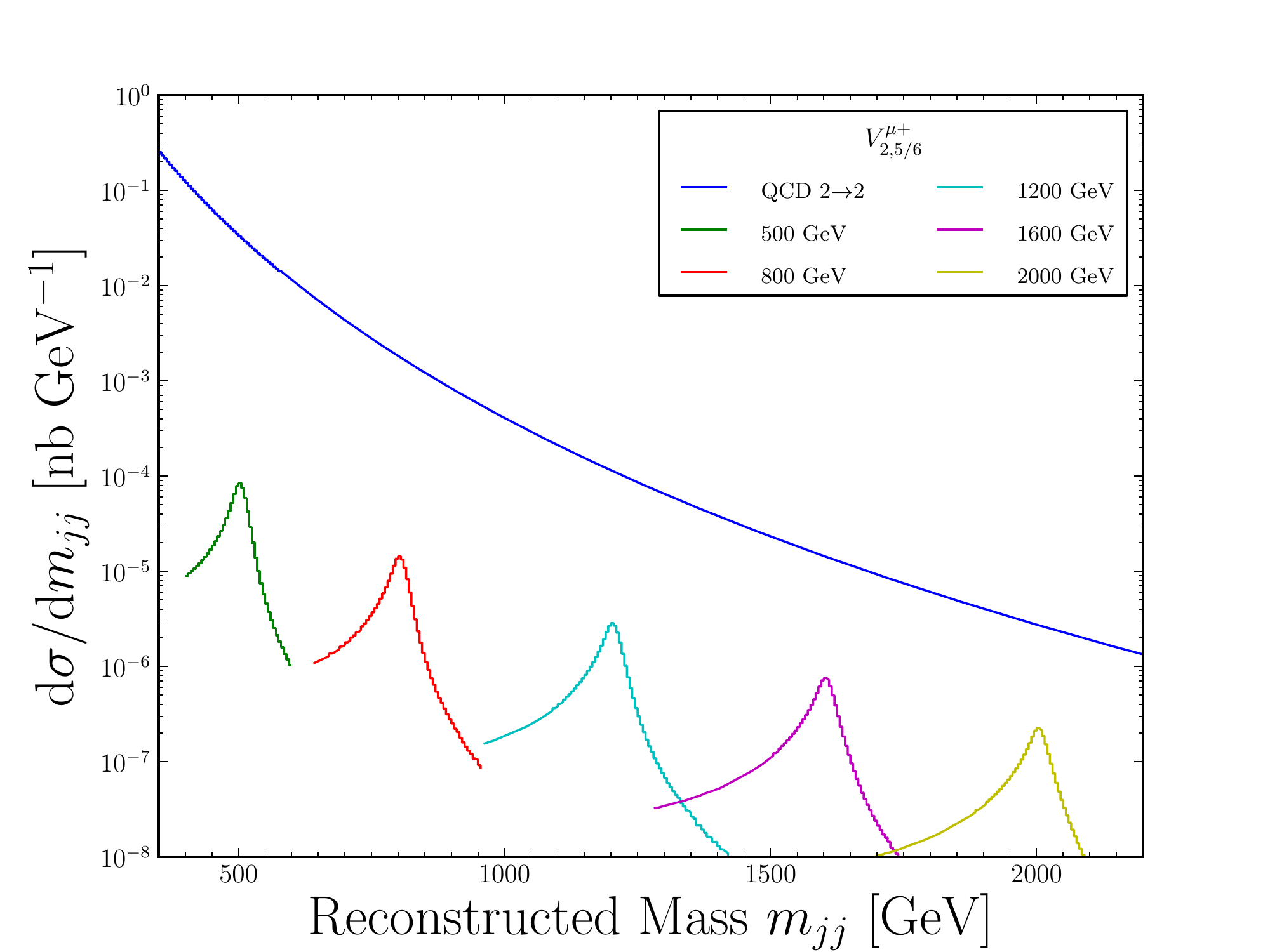}
  \label{subfig:diJetVectorMassA}
}
\end{center}
\caption{The dijet mass spectrum at $\sqrt{s} = \mathrm{7\,TeV}$ 
         for 500\,GeV, 800\,GeV, 1200\,GeV, 1600\,GeV and 2000\,GeV
         diquark masses with the couplings given in the text.}
\label{fig:diJetMassA}
\begin{center}
\subfigure[Scalar]{
  \includegraphics[width=0.47\textwidth]{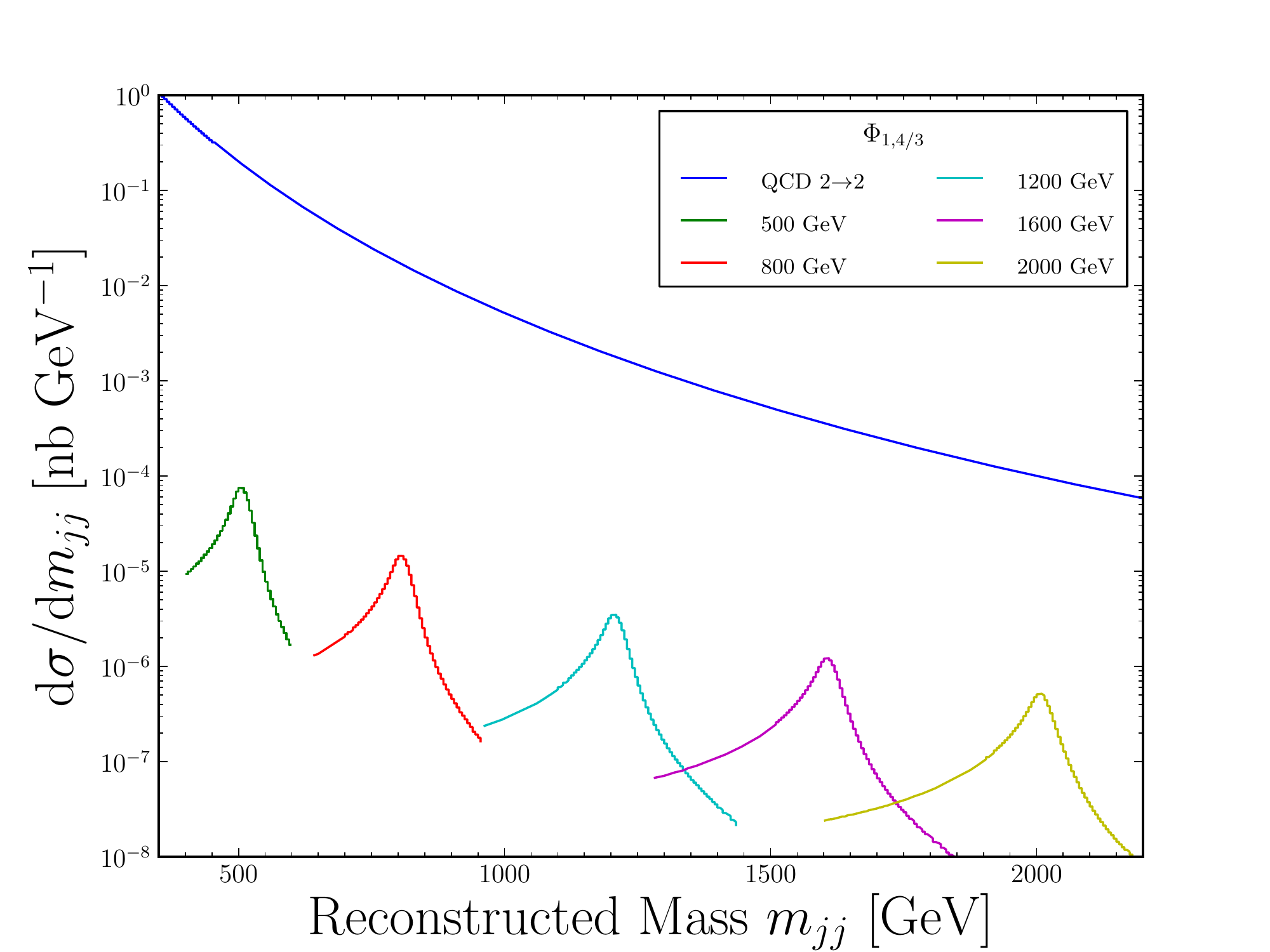}
  \label{subfig:diJetScalarMassB}
}
\subfigure[Vector]{ 
 \includegraphics[width=0.47\textwidth]{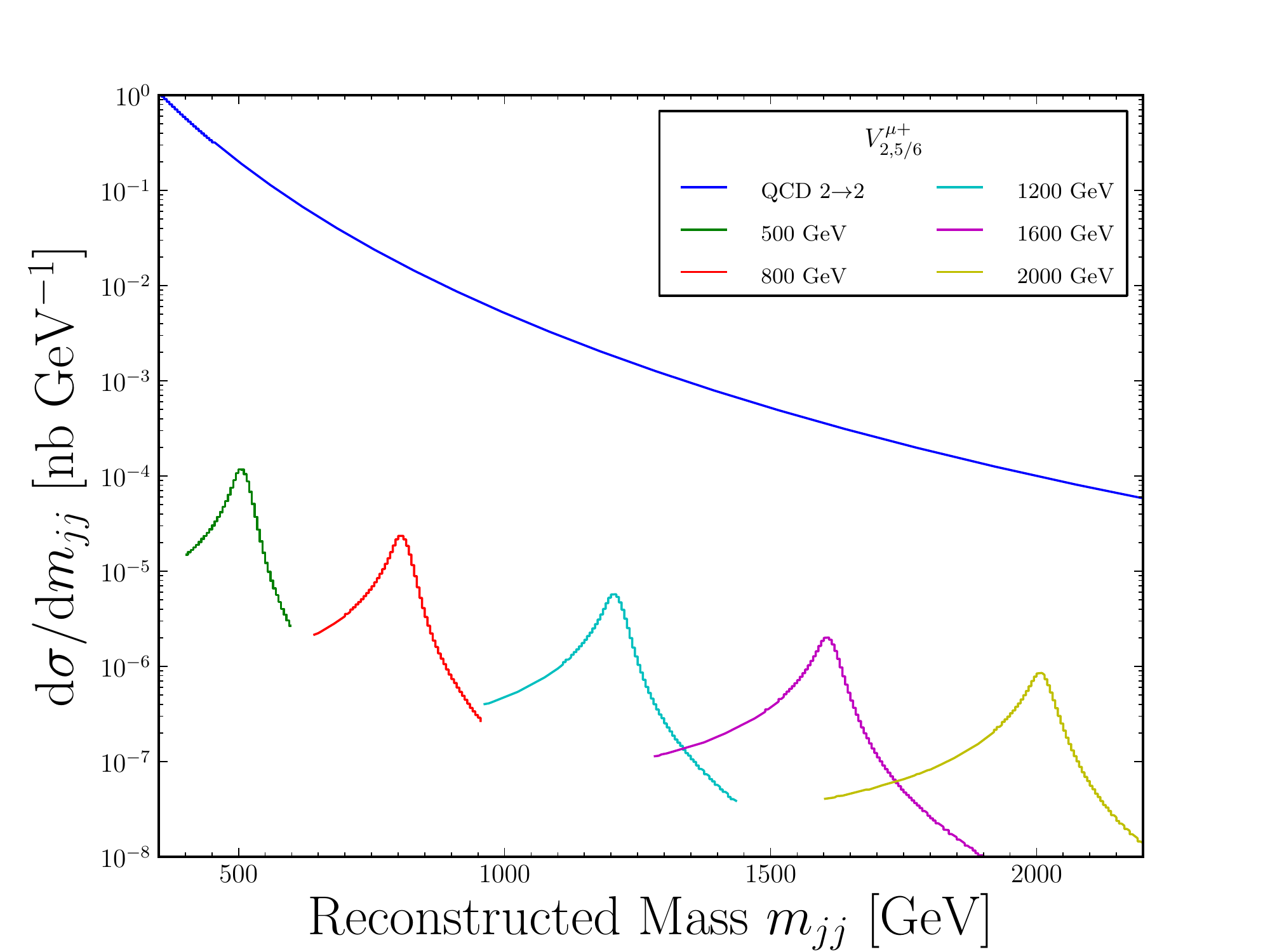}
  \label{subfig:diJetVectorMassB}
}
\end{center}
\caption{The dijet mass spectrum at $\sqrt{s} = \mathrm{14\,TeV}$ 
         for 500\,GeV, 800\,GeV, 1200\,GeV, 1600\,GeV and 2000\,GeV
         diquark masses with the couplings given in the text.}
\label{fig:diJetMassB}
\end{figure}

The signal and background were simulated using \textsf{Herwig++}.
The analysis and modelling of the backgrounds
followed that suggested in Ref.\,\cite{Aad:2011aj}.
The transverse momenta and pseudorapidities of the jets were required to
be  $p_T^1 > 150 \, \mathrm{GeV}$, $p_T^2 > 60 \,
\mathrm{GeV}$ and $|\eta_{1,2}| < 2.5$ where $1$ is the hardest jet and
$2$ is the subleading jet. In addition
we required that the dijet invariant mass, $m_{jj}$ satisfied \linebreak
\mbox{$m_{jj} > \mathrm{300\,GeV}$}
and the rapidity difference between the leading and subleading jet was
$|\Delta\eta_{12}| < 1.3$.
The dijet invariant mass spectrum after these cuts is 
shown in Figures~\ref{fig:diJetMassA}~and~\ref{fig:diJetMassB}
for $\sqrt s=7\ {\rm and}\ 14$\,TeV, respectively.
The diquarks were simulated at masses of 500\,GeV, 800\,GeV,
1200\,GeV, 1600\,GeV and 2000\,GeV.

As simulating the QCD $m_{jj}$ spectrum at high masses is difficult,
a functional form
\begin{equation}
f(x) = a_0(1-x)x^{(a_1+a_2\ln{x})},
\label{eq:massFit}
\end{equation} 
was fitted to the low masses,
where the $a_i$ are fitted parameters and $x=m_{jj}/\sqrt{s}$, and
extrapolated out into the high mass region.

The results of Ref.\,\cite{Aad:2011aj} can be used to impose
constraints on the diquark coupling as a function of the diquark mass.
The event selection from Ref.\,\cite{Aad:2011aj} was used to reproduce
the correct acceptance. This requires that the event 
contains at least two jets with $p_T > 150\ \mathrm{GeV}$
and a subleading jet with $p_T>30\ \mathrm{GeV}$.
Both the leading and subleading $p_T$ jets must
satisfy $|\eta_{j}| < 2.5$ with $\Delta\eta_{12} < 1.3$ and $m_{jj} >
150\ \mathrm{GeV}$. 

\begin{figure}[t]
\begin{center}
  \includegraphics[width=0.6\textwidth]{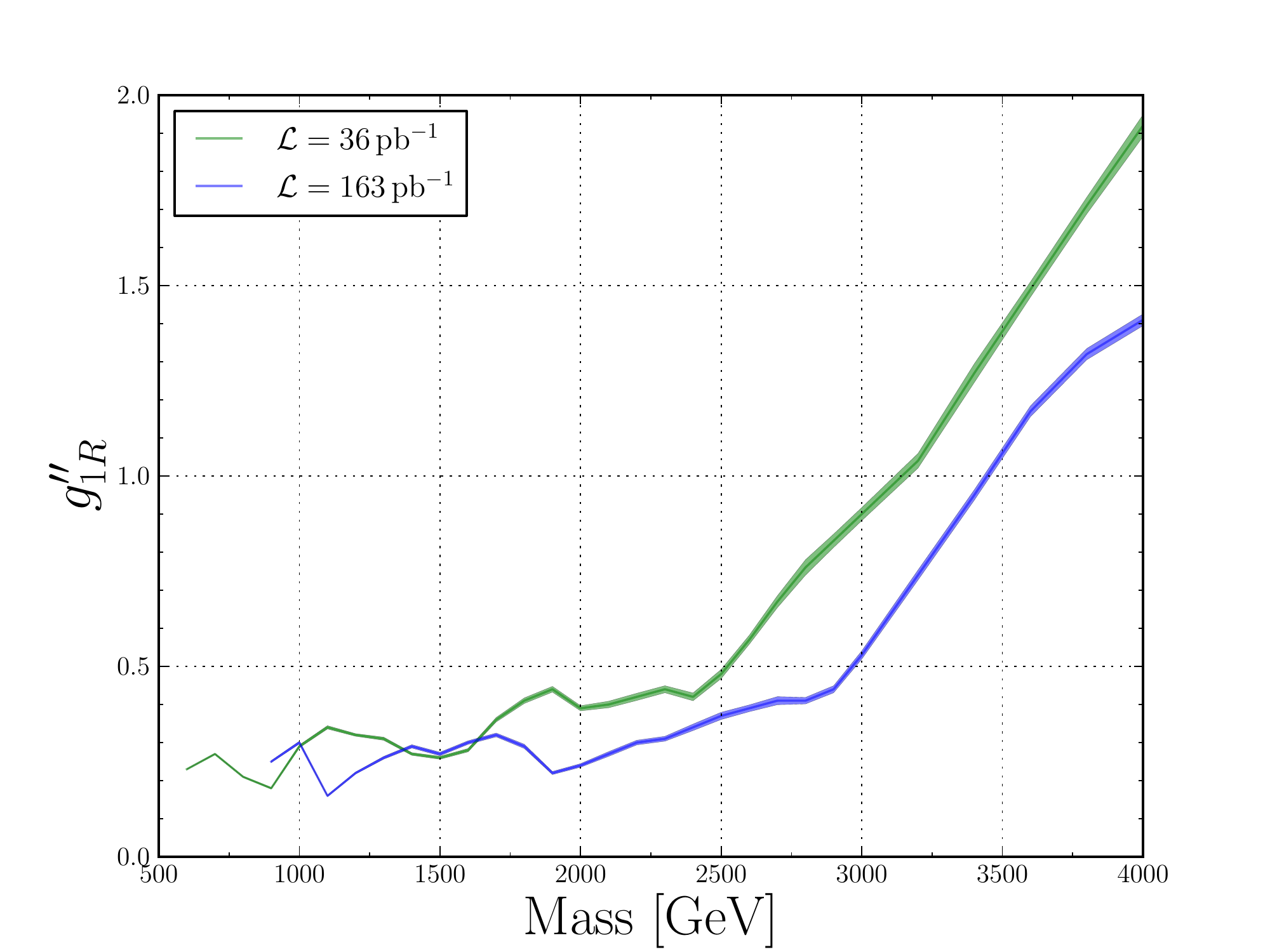}
\end{center}
\caption{Limit on the coupling as a function of the diquark mass based
         on the model independent data given in the recent ATLAS publications
         \cite{Aad:2011aj,ATLAS-CONF-2011-081}.  The band reflects the
         uncertainty from varying the scale between 50\% and 200\% of the diquark mass.}
\label{fig:couplingMassLimit}
\end{figure}

The signal, after the cuts, was fitted to a Gaussian distribution,
with the mean fixed, $m$, at the simulated diquark mass to obtain
the standard deviation of the distribution, $\sigma$, so that the results presented
in Ref.\,\cite{Aad:2011aj} could be used to obtain the limits on the diquark coupling.
As suggested in \cite{Aad:2011aj}, long tails were
removed by taking a window around the diquark mass of $\pm 20 \%$ for
the fit. If the $\sigma/m$ value obtained was below the
range of that given in the paper, then the number of events associated
with the lowest $\sigma/m$ for that mass was used. This allows a
conservative estimate for the excluded coupling, as opposed to one
which may be obtained by extrapolation into the unknown region.
The limit on the diquark coupling is shown in Figure~\ref{fig:couplingMassLimit}
where because the statistical errors were negligible, the bands shown come
from varying the scale from $50 \%$ to $200 \%$  of the default scale choice,
{\it i.e.} the diquark mass.

Following the work of Ref.\,\cite{Aad:2011aj} the ATLAS collaboration
has released an updated analysis~\cite{ATLAS-CONF-2011-081}, including
additional data corresponding to an integrated luminosity of $163\,{\rm pb}^{-1}$.
This analysis included slightly harder cuts requiring $p_T > 180\ \mathrm{GeV}$ and
$m_{jj} > 170\ \mathrm{GeV}$ in addition to the cuts used in Ref.\,\cite{Aad:2011aj}.
The limit obtained from this higher integrated luminosity analysis is also
shown in Figure~\ref{fig:couplingMassLimit}.
We note that ATLAS performed better than the expected median
limit in the Ref.\,\cite{Aad:2011aj} and worse than the expected median
limit in \cite{ATLAS-CONF-2011-081} in the $1400 - 1600 \, \mathrm{GeV}$
mass range, giving rise to the overlap in Figure~\ref{fig:couplingMassLimit}.

\subsection{Pair Production}

\begin{figure}[t]
  \begin{center}
    \subfigure[$s$-channel]{
      \includegraphics[width=0.4\textwidth]{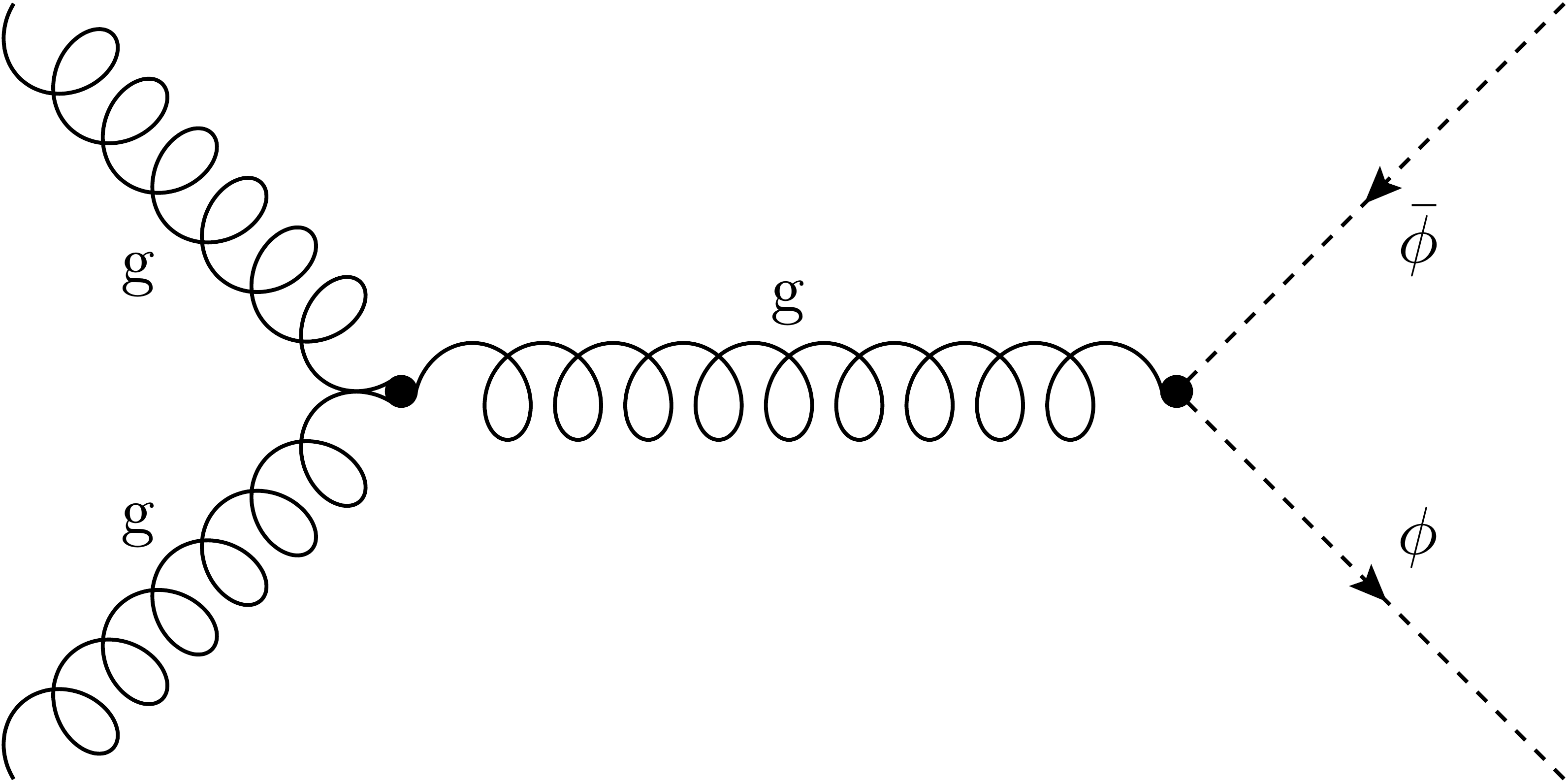}
      \label{subfig:pairSChannel}
    }
    \subfigure[$t$-channel]{ 
      \includegraphics[width=0.4\textwidth]{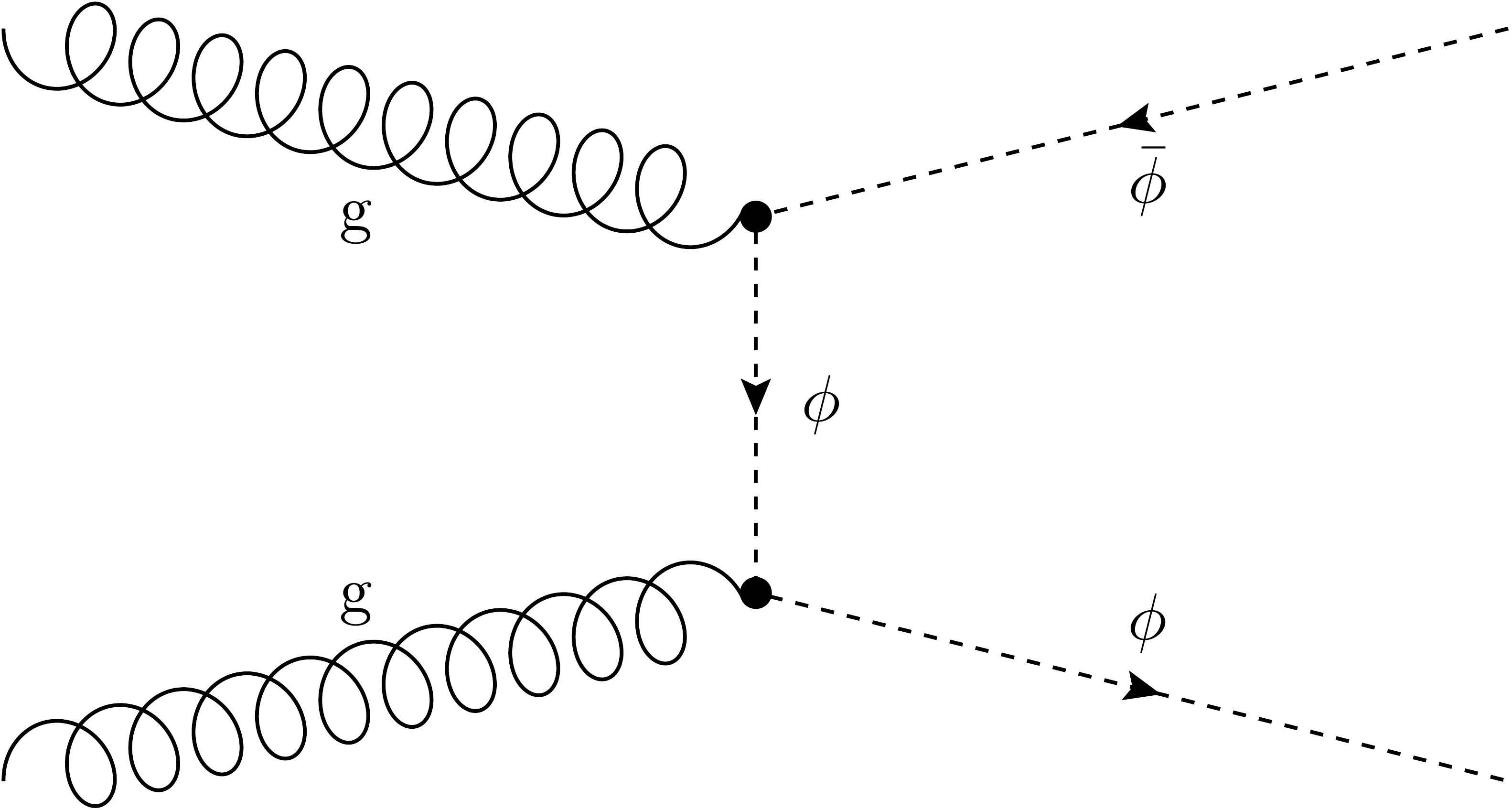}
      \label{subfig:pairTChannel}
    } 
    \\
    \subfigure[$u$-channel]{ 
      \includegraphics[width=0.4\textwidth]{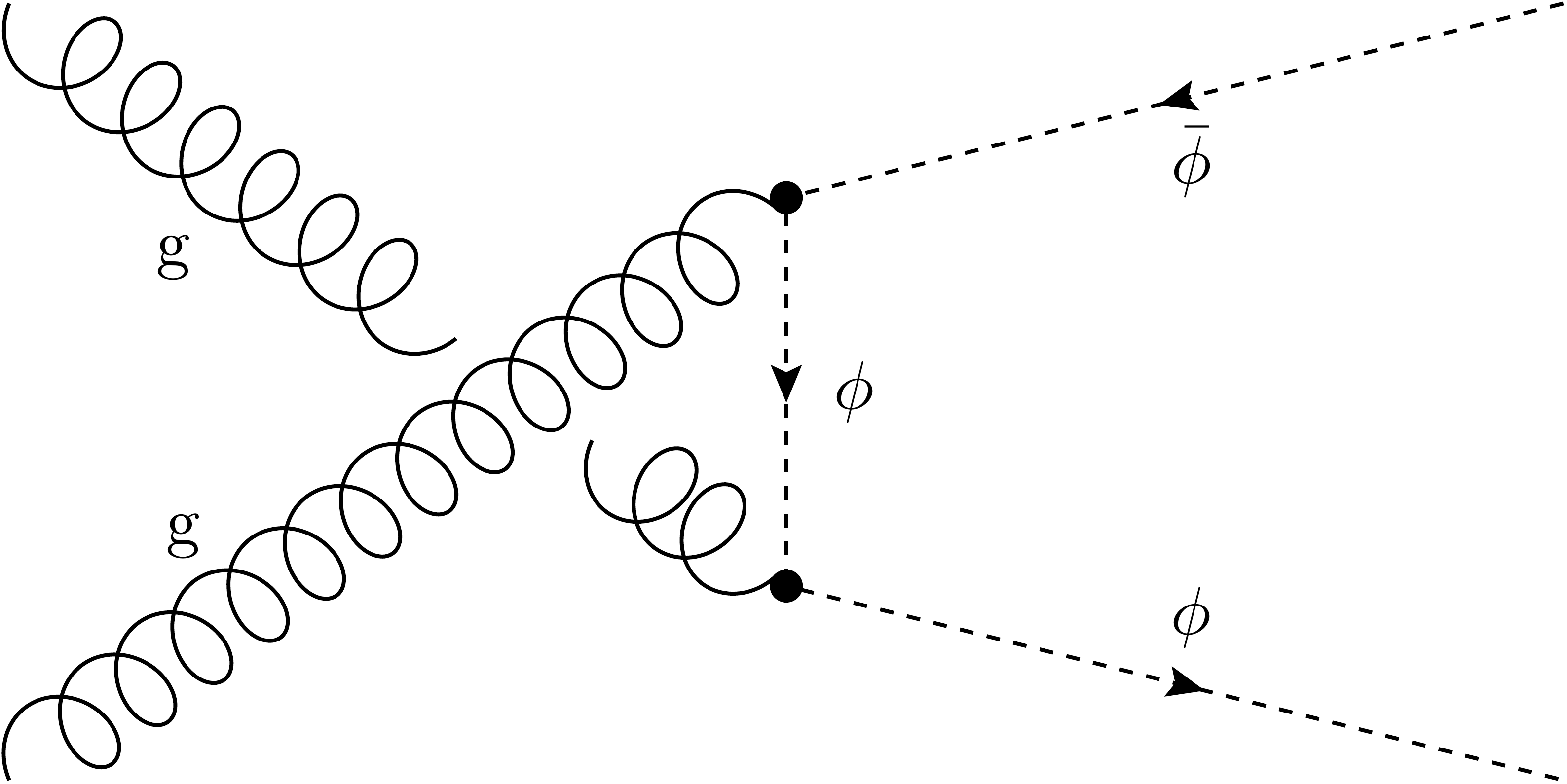}
      \label{subfig:pairUChannel}
    }
    \subfigure[4-point Channel]{ 
      \includegraphics[width=0.4\textwidth]{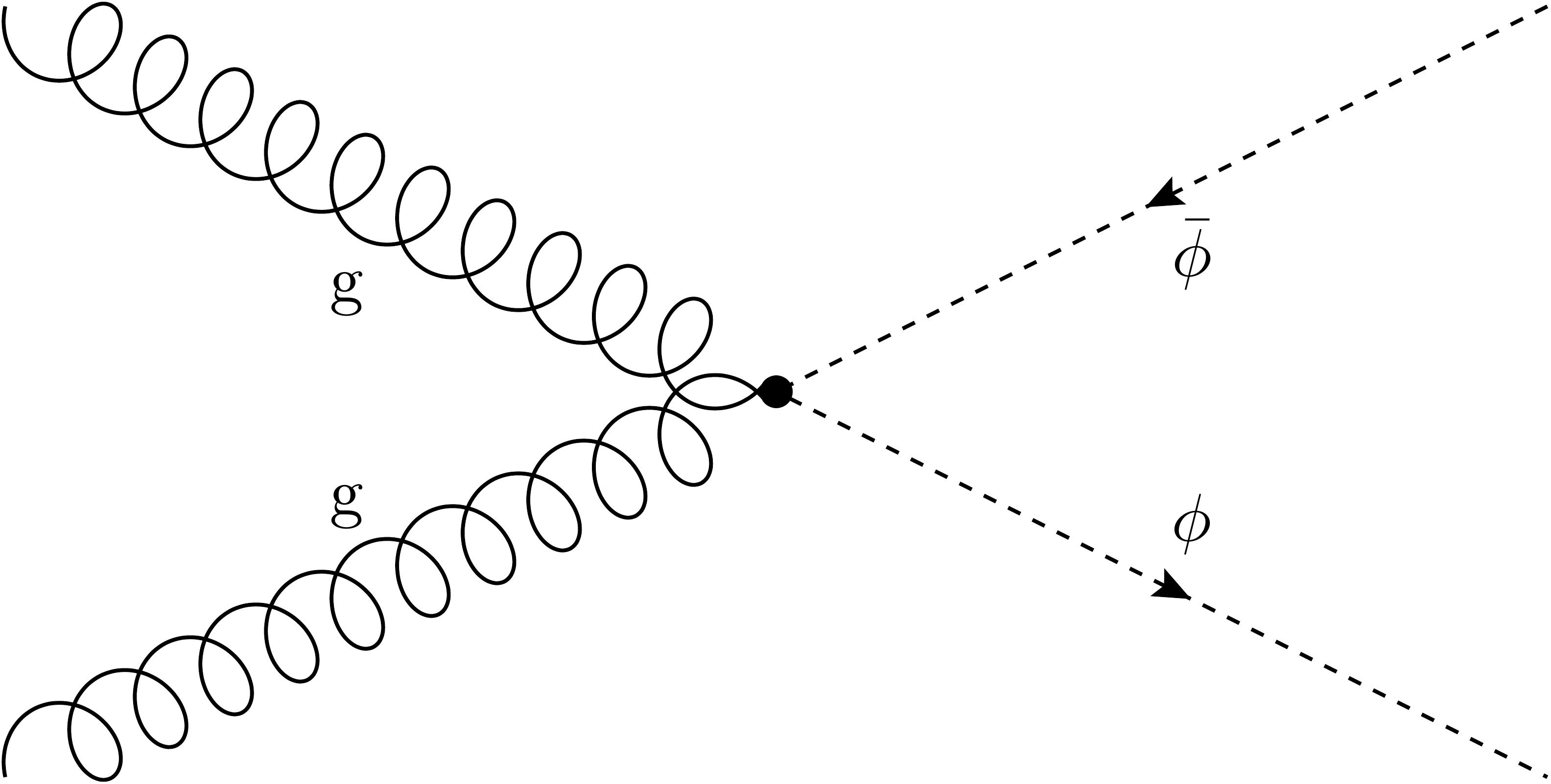}
      \label{subfig:pairFourPointChannel}
    }
    \caption{The four diagrams contributing to the pair production of a diquark.}
    \label{fig:pairProduction}
  \end{center}
\end{figure}

The pair production of diquarks (scalar and vector) occurs via the 
Feynman diagrams shown in Figure~\ref{fig:pairProduction}.
The pair production process has
one main advantage over the resonant production, it does not depend on
the unknown diquark coupling. Instead, the pair production cross section
depends only on the $SU(3)_C$ representation, mass and spin of the particle.
The pair production processes therefore has the potential
to distinguish between whether a particle in the antitriplet
or sextet representation was produced due to the dependency
of the cross section on the colour representation.

To date there have been no studies of the experimental signals of
diquark pair production.  The cross section has been calculated
\cite{Tanaka:1991nr,Chen:2008hh,Patel:2011eh} and some work towards a jet
study, no Monte Carlo study has been performed.

The pair production and subsequent decay of diquarks is expected to
give four jets, with two pairs of jets forming systems with
the mass of the diquark. The backgrounds to the pair production of
diquarks are:
\begin{itemize}
\item vector boson $WW$, $ZZ$ and $ZW$ pair production;
\item vector boson, $W$ and $Z$,  production in association with additional
      jets;
\item top quark pair, $t\bar{t}$, production;
\item QCD jet production.
\end{itemize}

The analysis proceeded by placing cuts on the four hardest jets: $p_T^1
> 150 \, \mathrm{GeV}$, $p_T^2 > 100 \, \mathrm{GeV}$, $p_T^3 > 60 \,
\mathrm{GeV}$ and $p_T^4 > 30 \, \mathrm{GeV}$, where the four 
jets $i=1,4$ are ordered in $p_T$ such that the first jet is the hardest.
All four jets were required to have pseudorapidity $|\eta_{i}| < 3$.
Two pairs of jets were then formed with the pairing selected
that minimized the mass
difference between the two pairs of jets. 
 If, after pairing, the two hardest
jets are in the same pair of jets, the event was vetoed.
  The mass
difference between the pairs was required to be less than $20 \,
\mathrm{GeV}$.

\begin{figure}[t!!]
\begin{center}
\subfigure[Scalar]{
  \includegraphics[width=0.47\textwidth]{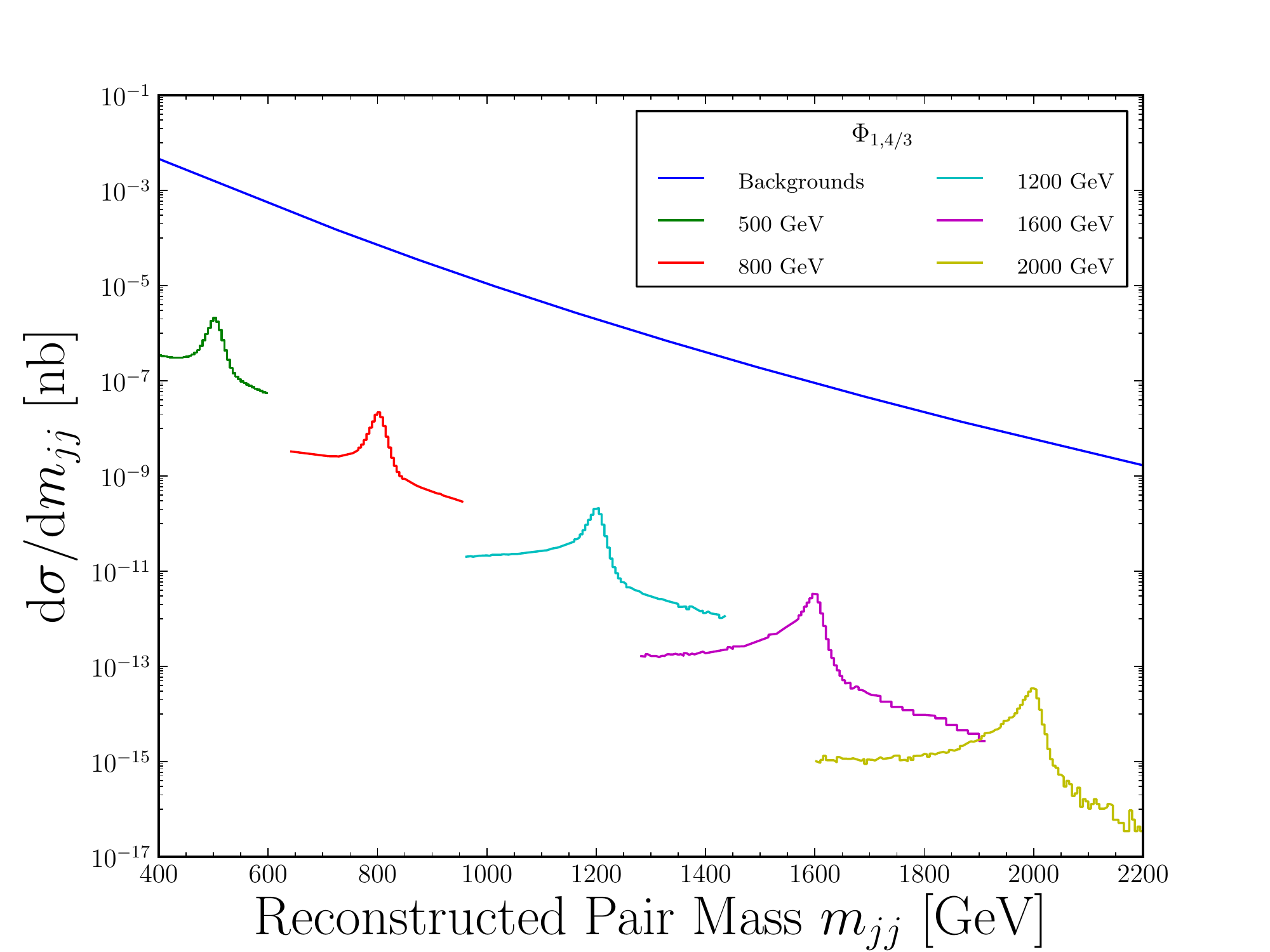}
  \label{subfig:fourJetScalarMassA}
}
\hfill
\subfigure[Vector]{ 
 \includegraphics[width=0.47\textwidth]{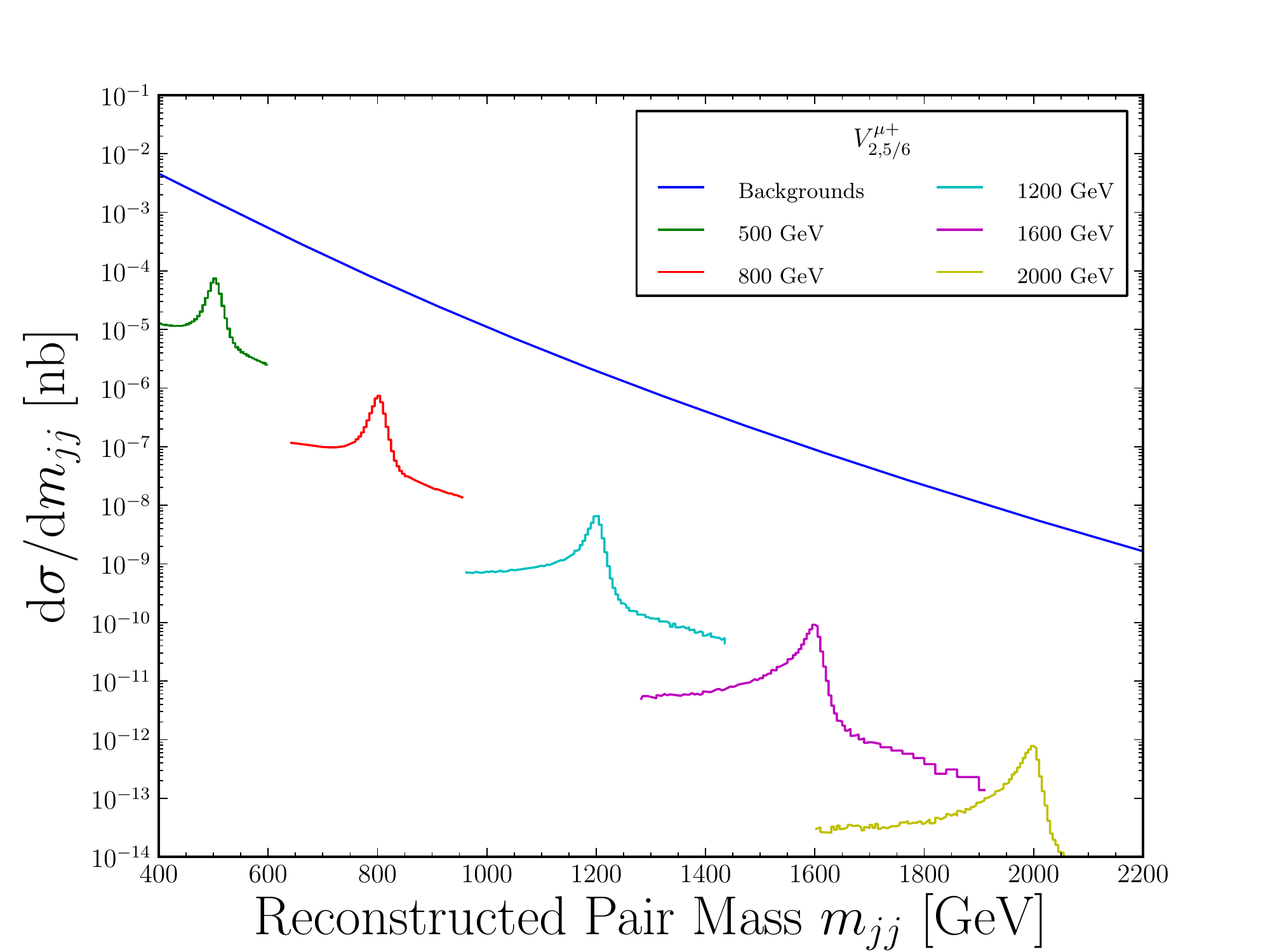}
  \label{subfig:fourJetVectorMassA}
}
\end{center}
\caption{The mass spectrum of dijet pairs in four jet events at 
         $\sqrt{s} = \mathrm{7\,TeV}$ for 500\,GeV, 800\,GeV, 1200\,GeV, 1600\,GeV
         and 2000\,GeV diquark masses.}
\label{fig:fourJetMassA}
\begin{center}
\subfigure[Scalar]{
  \includegraphics[width=0.47\textwidth]{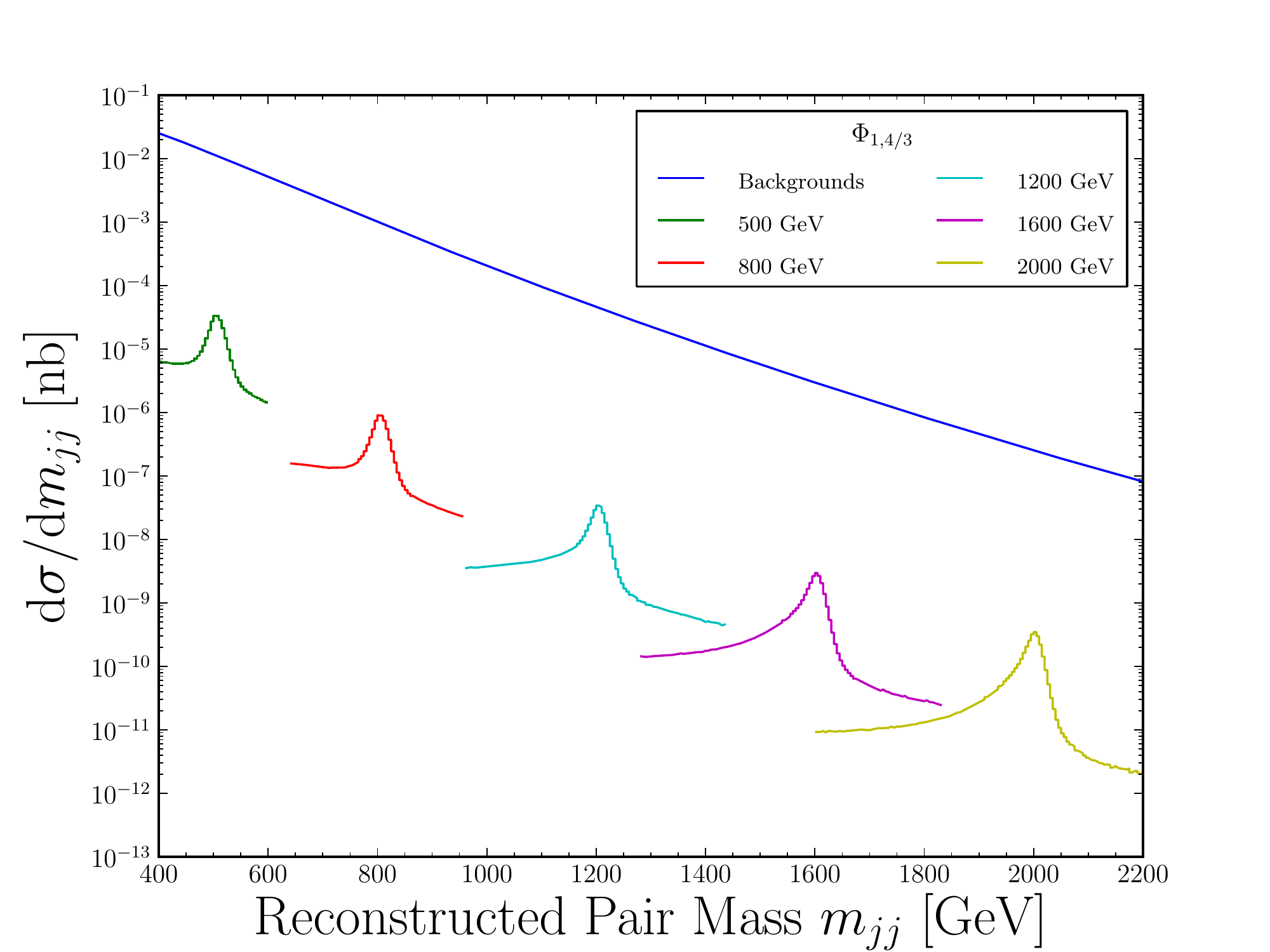}
  \label{subfig:fourJetScalarMassB}
}\hfill
\subfigure[Vector]{ 
 \includegraphics[width=0.47\textwidth]{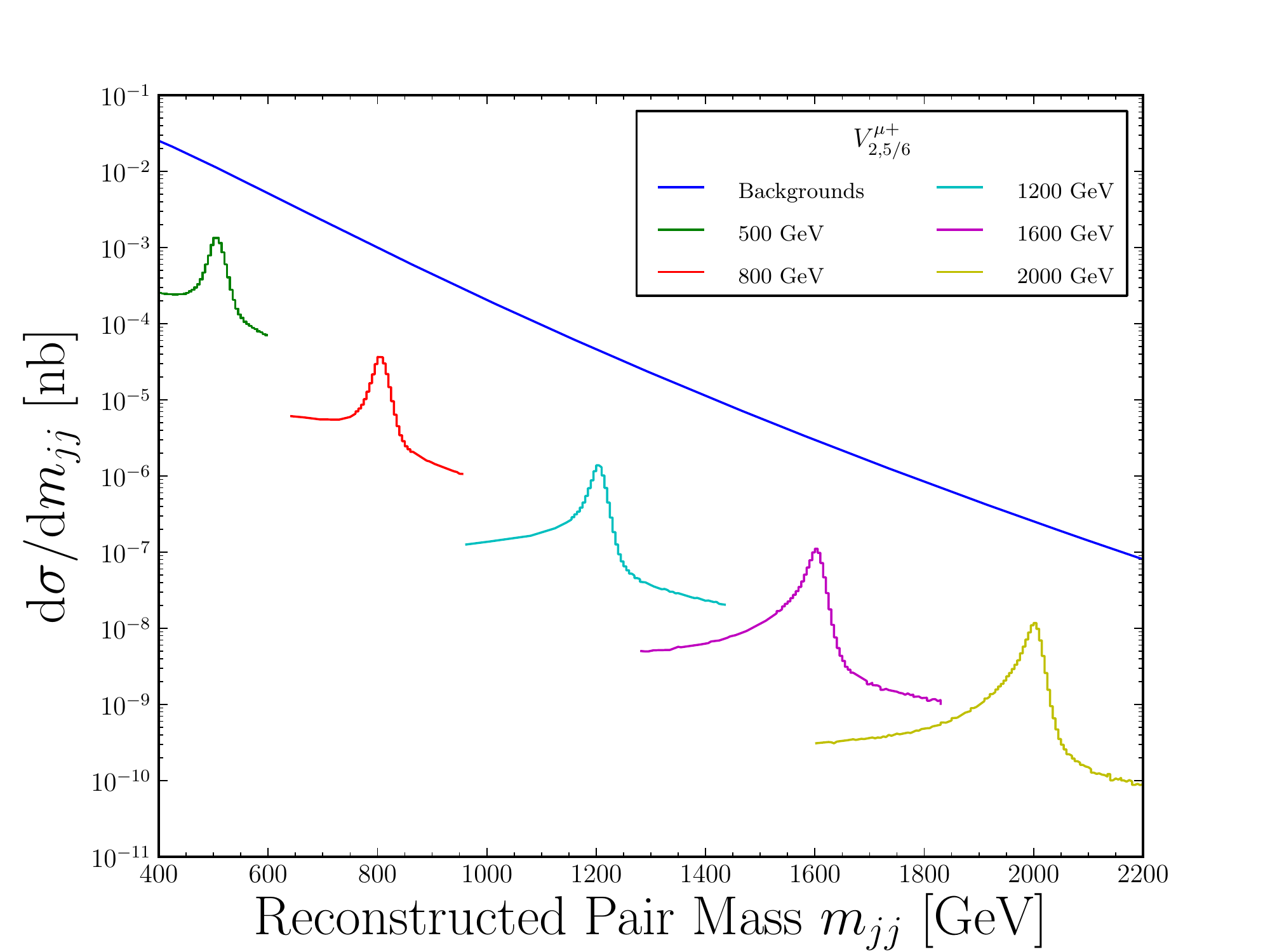}
  \label{subfig:fourJetVectorMassB}
}
\end{center}
\caption{The mass spectrum of dijet pairs in four jet events at 
         $\sqrt{s} = \mathrm{14\,TeV}$ for 500\,GeV, 800\,GeV, 1200\,GeV, 1600\,GeV
         and 2000\,GeV diquark masses.}
\label{fig:fourJetMassB}
\end{figure}

The signal and backgrounds were simulated for the production of the
$\Phi_{1,1/3}$ and $V^{\mu +}_{2,5/6}$ diquarks giving the results
shown in Figures~\ref{fig:fourJetMassA}~and~\ref{fig:fourJetMassB} for
\linebreak \mbox{$\sqrt s=7\ {\rm and}\ 14$\,TeV}, respectively.  As
the backgrounds are dominated by QCD scattering, {\it i.e.} the
contribution of the QCD scattering processes is approximately one
hundred times that of all the other backgrounds combined, only the sum
of the backgrounds is shown. As for the resonant production, the plots
show the production of diquarks with masses of 500\,GeV, 800\,GeV and
1200\,GeV, 1600\,GeV and 2000\,GeV.  The low mass QCD background was
fitted with Eq.~\ref{eq:massFit} and extended out into the high mass
region.

\begin{figure}[t!!]
  \begin{center}
    \subfigure[7 TeV]{
      \includegraphics[width=0.47\textwidth]{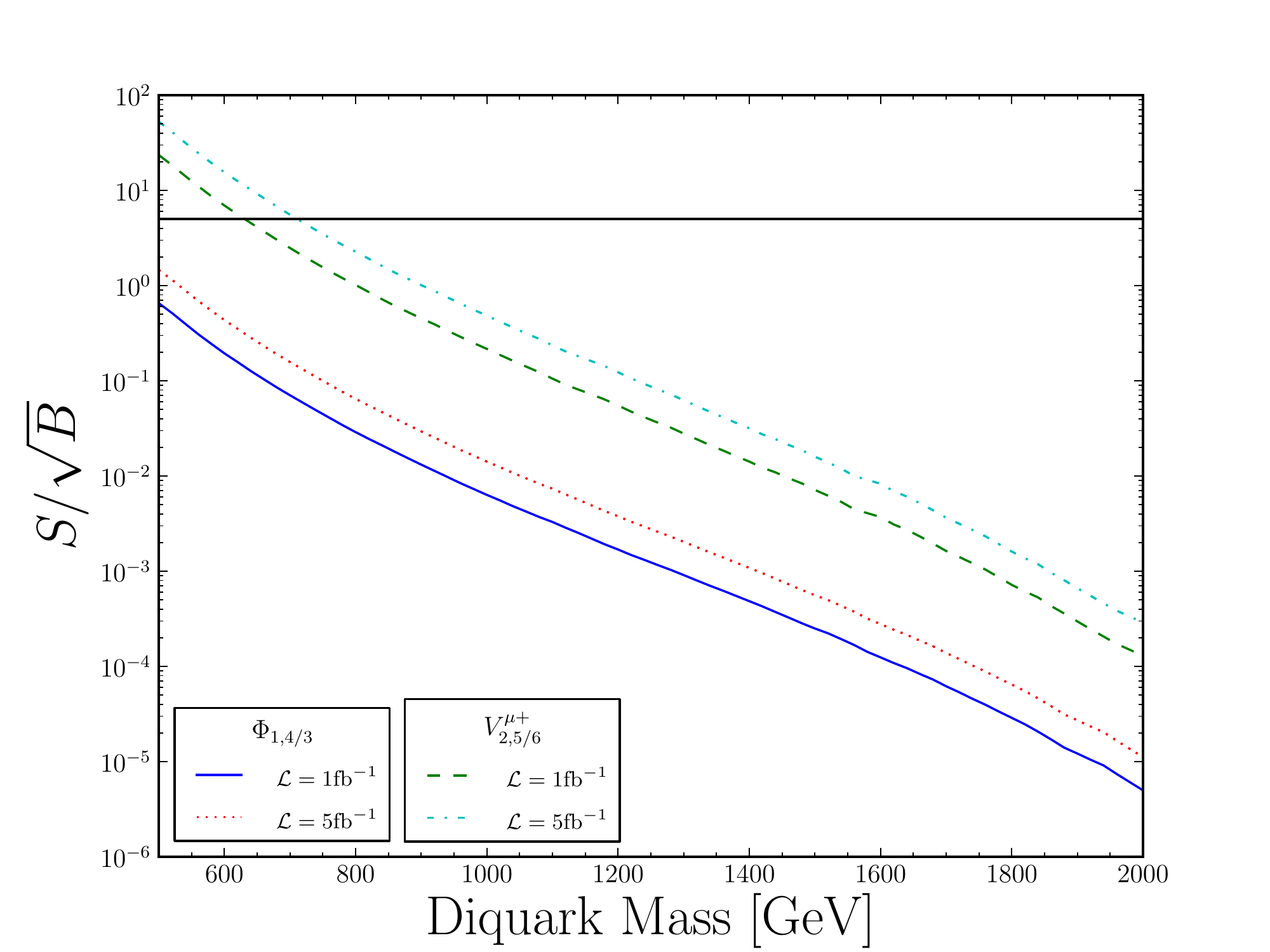}
      \label{subfig:sig7TeV}
    }\hfill
    \subfigure[14 TeV]{
      \includegraphics[width=0.47\textwidth]{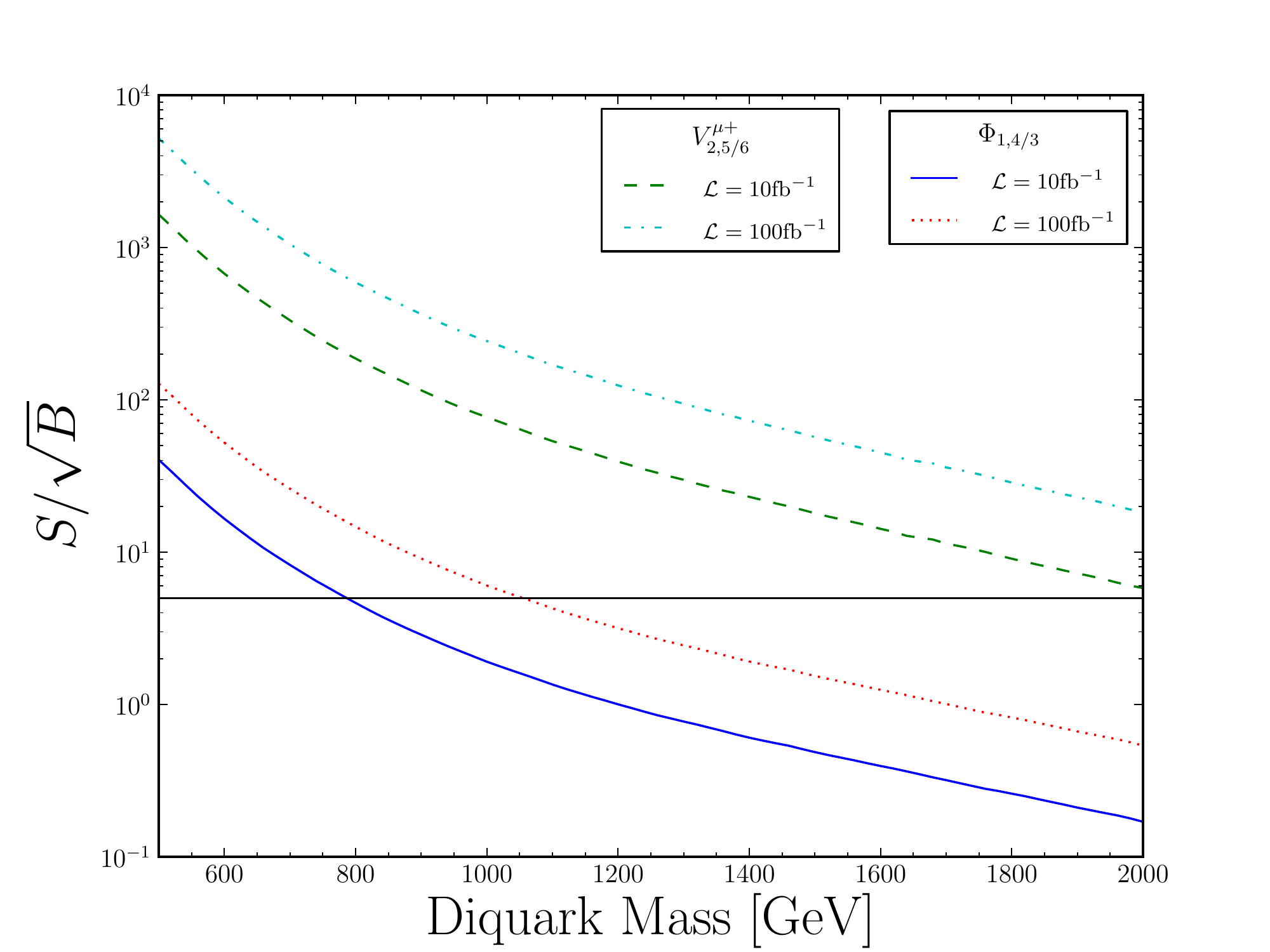}
      \label{subfig:sig14TeV}
    }
  \end{center}
  \caption{$\tfrac{S}{\sqrt{B}}$ for the scalar and vector diquark at
    luminosities of $\mathcal{L} = \mathrm{1\, fb^{-1}}$ and
    $\mathcal{L} = \mathrm{5\, fb^{-1}}$ at $\sqrt{s} = \mathrm{7\,TeV}$
    and $\mathcal{L} = \mathrm{10\, fb^{-1}}$ and $\mathcal{L} =
    \mathrm{100\, fb^{-1}}$ at\protect\linebreak
    \mbox{$\sqrt{s} = \mathrm{14\,TeV}$}. The black
    horizontal line shows $\frac{S}{\sqrt{B}} = 5$.}
  \label{fig:sig}
\end{figure}

A window of $\pm \mathrm{50\, GeV}$ was taken around the diquark mass
and the $\tfrac{S}{\sqrt{B}}$ was calculated for a number of luminosities the
results of which are shown in Figure~\ref{fig:sig}. We see that that
the vector diquark manifests itself more prominently than the scalar,
which is consistent with the increased cross section of the vector in
the pair production process as seen in Figure~\ref{fig:crossSection}.
It will be hard to observe a scalar diquark using the pair production
process at $\sqrt{s}=7$\,TeV while with $\mathcal{L}=5\,\rm{fb}^{-1}$
it should be possible to observe a vector diquark with mass less than $700$\,GeV.

There is a marked increase in discovery potential at the increased
energy and luminosities running at $\sqrt{s} =\mathrm{14\,TeV}$
brings. A vector diquark in the mass range presented
here~($<2000$\,GeV) should be seen with
$\mathcal{L}=10\,\rm{fb}^{-1}$, whereas even with
$\mathcal{L}=100\,\rm{fb}^{-1}$ only a low mass~($\lesssim1050$\,GeV)
scalar diquark has potential for discovery.

\section{Conclusions}

In this paper we have presented a method for simulating the production and decay
of particles in the sextet colour representation. This approach
has been implemented in \textsf{Herwig++} and will be available in a forthcoming release.

This new simulation was used to simulate the production and decay of
vector and scalar sextet diquarks at energies relevant to the LHC.
Based on the findings and the latest ATLAS search
for new particles in two-jet final states, new constraints have been
put on the couplings of the diquarks to SM particles.

We have presented the first studies of the pair production mechanism, which
is independent of the unknown coupling of the sextet diquark to quarks.
This process has a promising search reach with the possibility of
observing vector diquarks with masses less than 710\,GeV at $\sqrt{s}=7$\,TeV
and both vector, for masses less than 2\,TeV, and scalar, for masses less
than 1\,TeV, with the LHC running at design energy. Hopefully the availability
of a Monte Carlo simulation of these processes will allow a more detailed 
experimental study.

\section{Acknowledgements}

We are grateful to all the other members of the \textsf{Herwig++}
collaboration for valuable discussions. We acknowledge the use of the
UK Grid for Particle Physics in producing the results. This work was
supported by the Science and Technology Facilities Council.
DW acknowledges support by the STFC
studentship ST/F007299/1.

\appendix
\label{app:all}
\section{Colour Decomposition}

\textsf{Herwig++}, as with all general purpose Monte Carlo generators,
has all the machinery set up to work with the (anti)fundamental
representation of $SU(3)_C$. Therefore exotic color
representations must be decomposed into a fundamental representation
basis.

The results of the this decomposition for the resonant and pair
production of sextet diquarks and the radiation of a gluon from a
sextet diquark are explicitly outlined below.

\subsection{Resonant Production}
\begin{figure}[h]
  \begin{center}
    \subfigure[]{
      \includegraphics[width=0.35\textwidth]{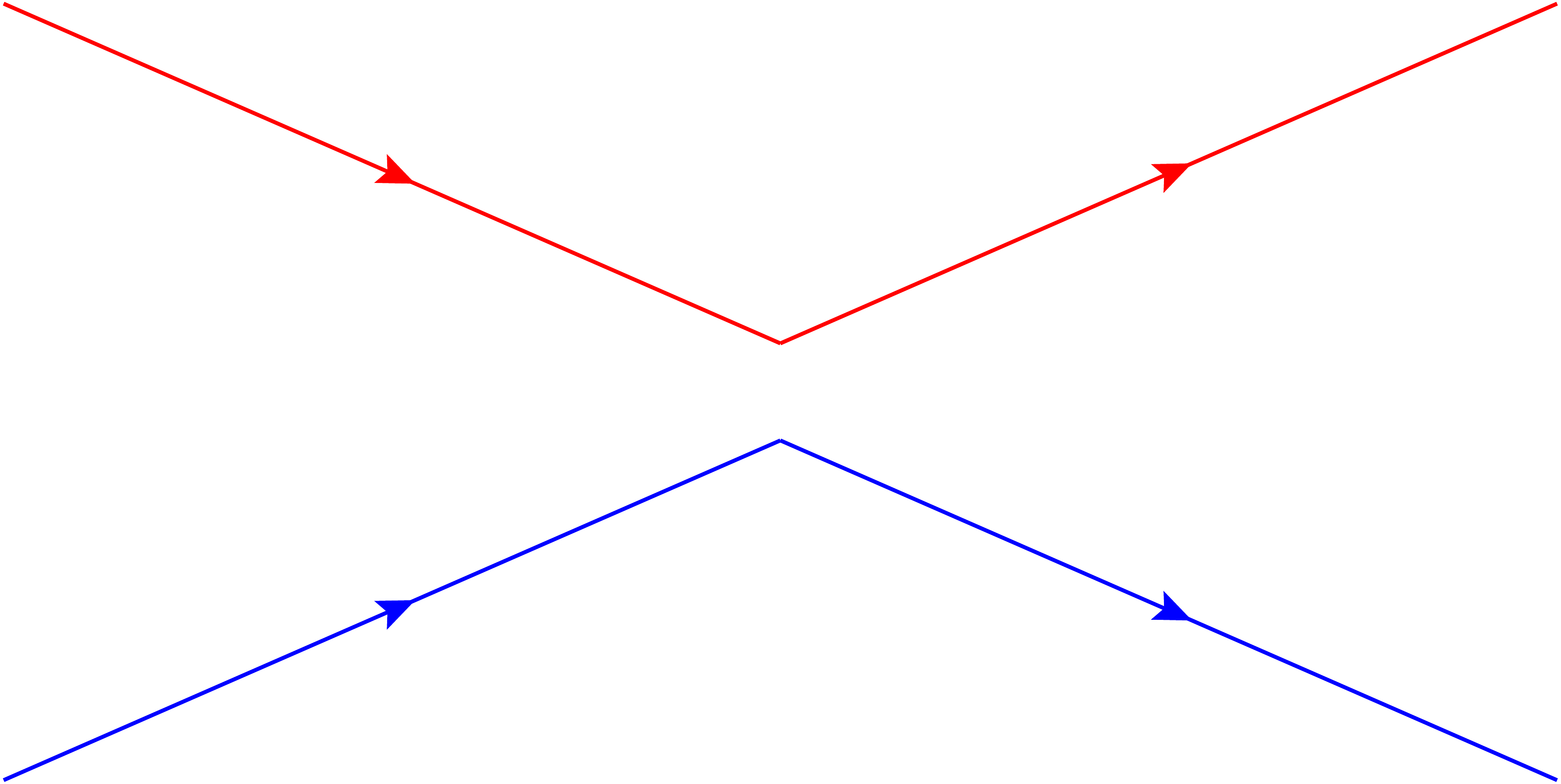}
    }\hfill
    \subfigure[]{
      \includegraphics[width=0.35\textwidth]{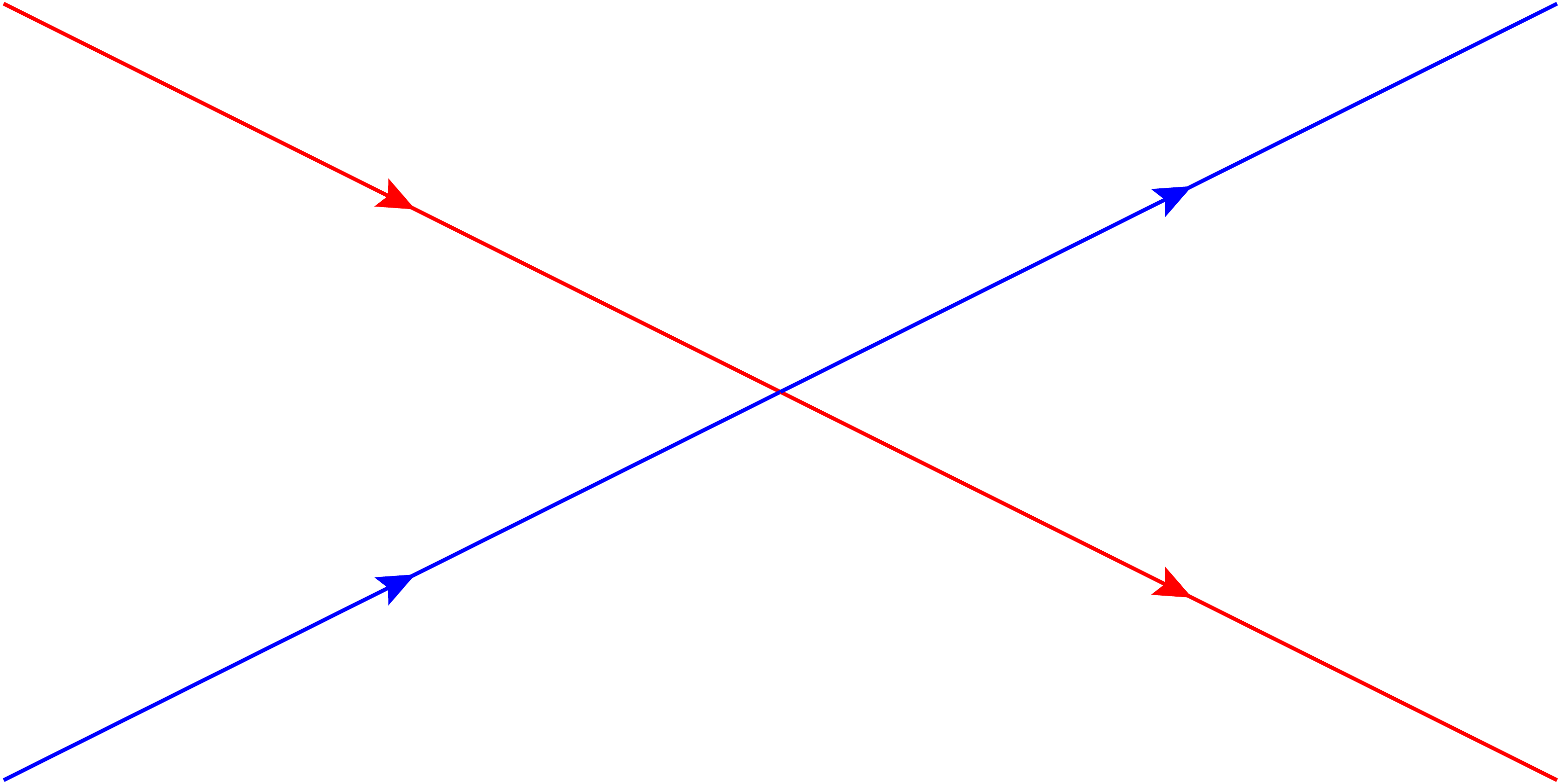}
    }
    \caption{Unique colour flows associated with the 
      resonant production of a diquark.}
    \label{fig:resColourFlows}
  \end{center}
\end{figure}

The diagram for resonant production and decay is shown in
Figure~\ref{fig:resProduction}. Following decomposition into the
fundamental representation, there are two unique colour flows
associated with this process as shown in
Figure~\ref{fig:resColourFlows}.  The colour factor associated with
these colour flows is $\tfrac{N_C(N_C+1)}{2}$.
\\

\subsection{Pair Production}
The diagrams contributing to the pair production of diquarks are shown
in Figure~\ref{fig:pairProduction}. Each of these diagrams is
decomposed into a colour factor and colourless component. By taking
the colour factors for the diagrams, adding and squaring, a table of
colour factors can be produced for each term in the total matrix
element squared.

\begin{figure}[]
  \begin{center}
    \subfigure[]{
      \includegraphics[width=0.27\textwidth]{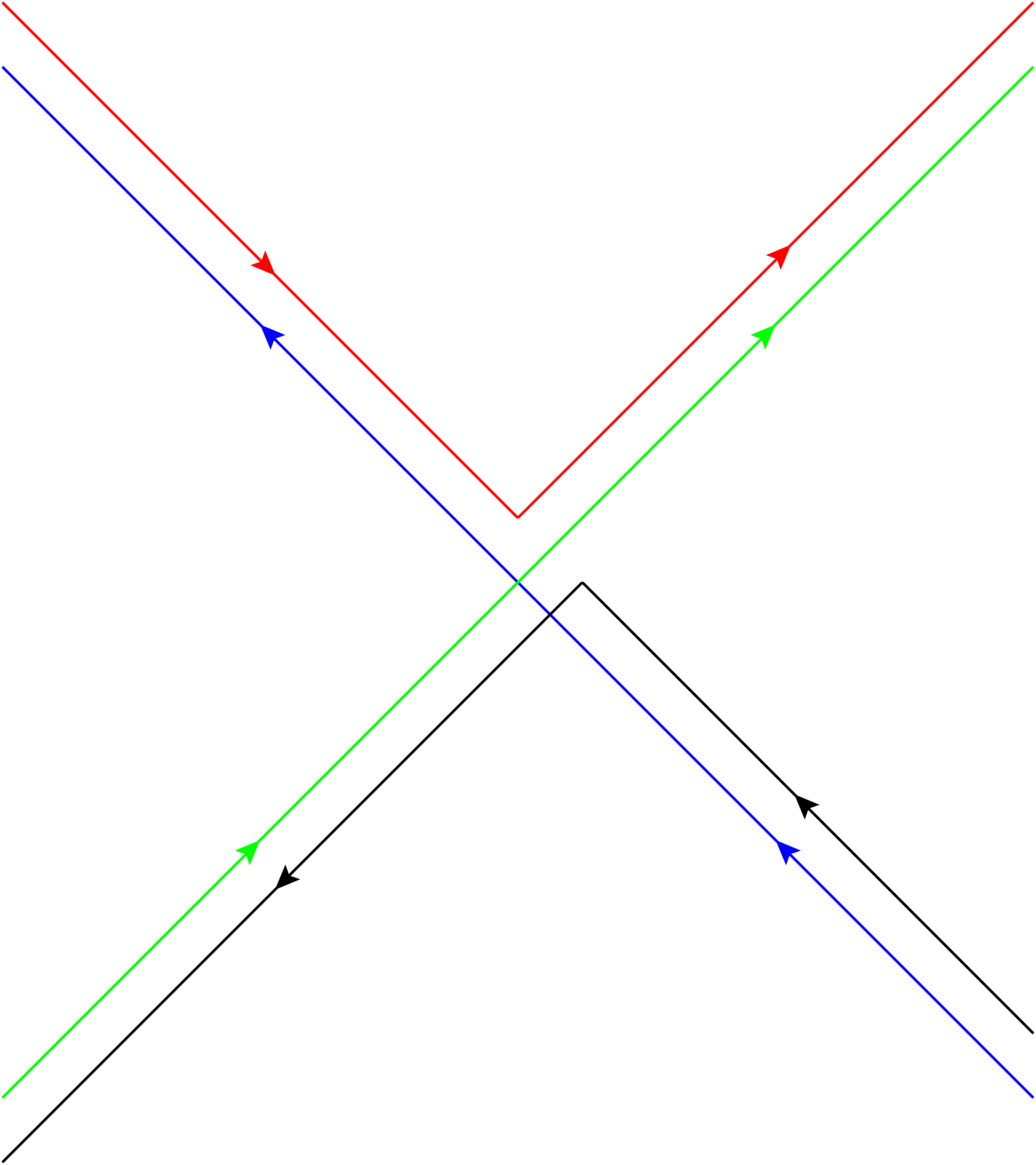}
    }
    \subfigure[]{
      \includegraphics[width=0.27\textwidth]{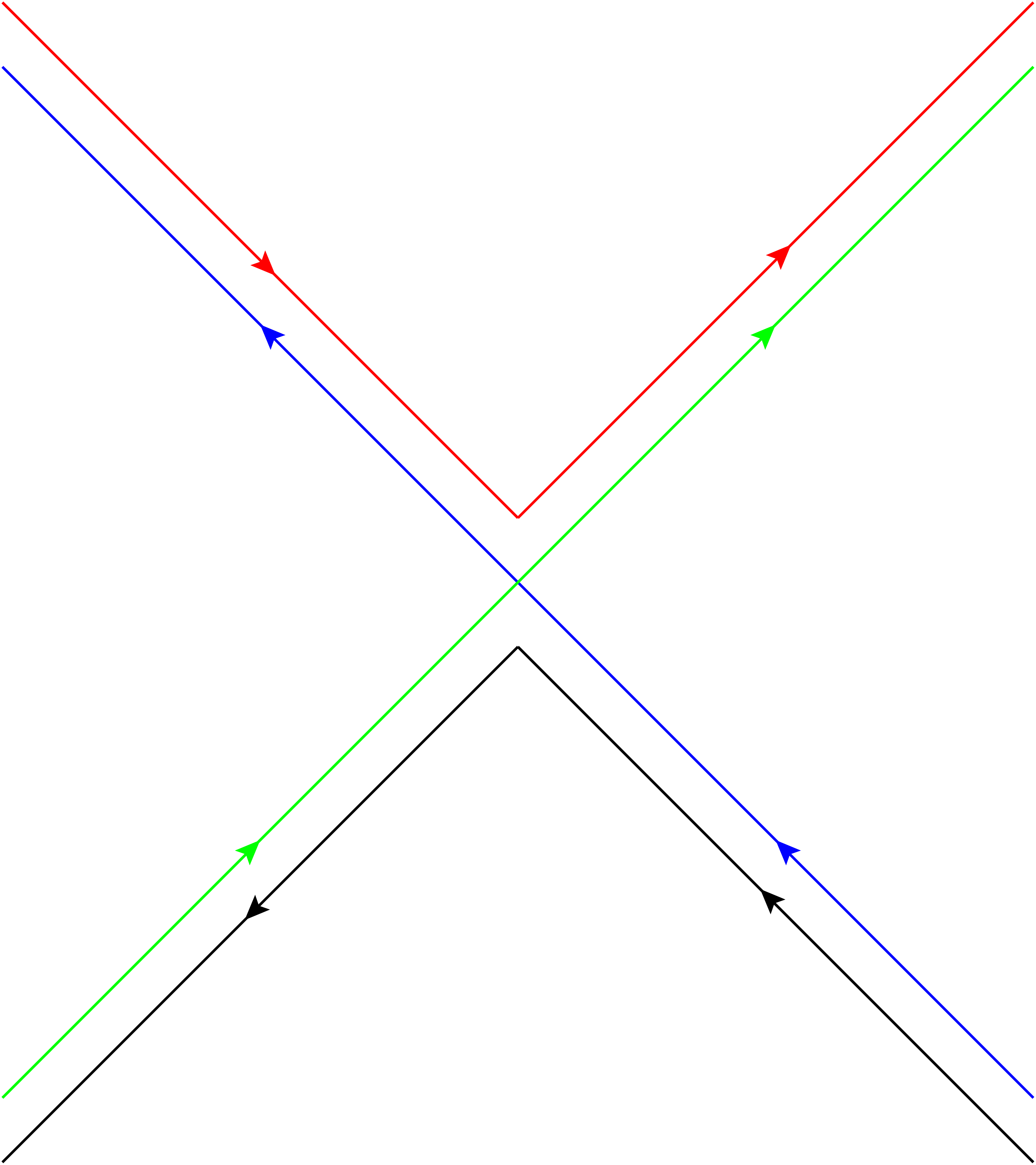}
    }
    \subfigure[]{
      \includegraphics[width=0.27\textwidth]{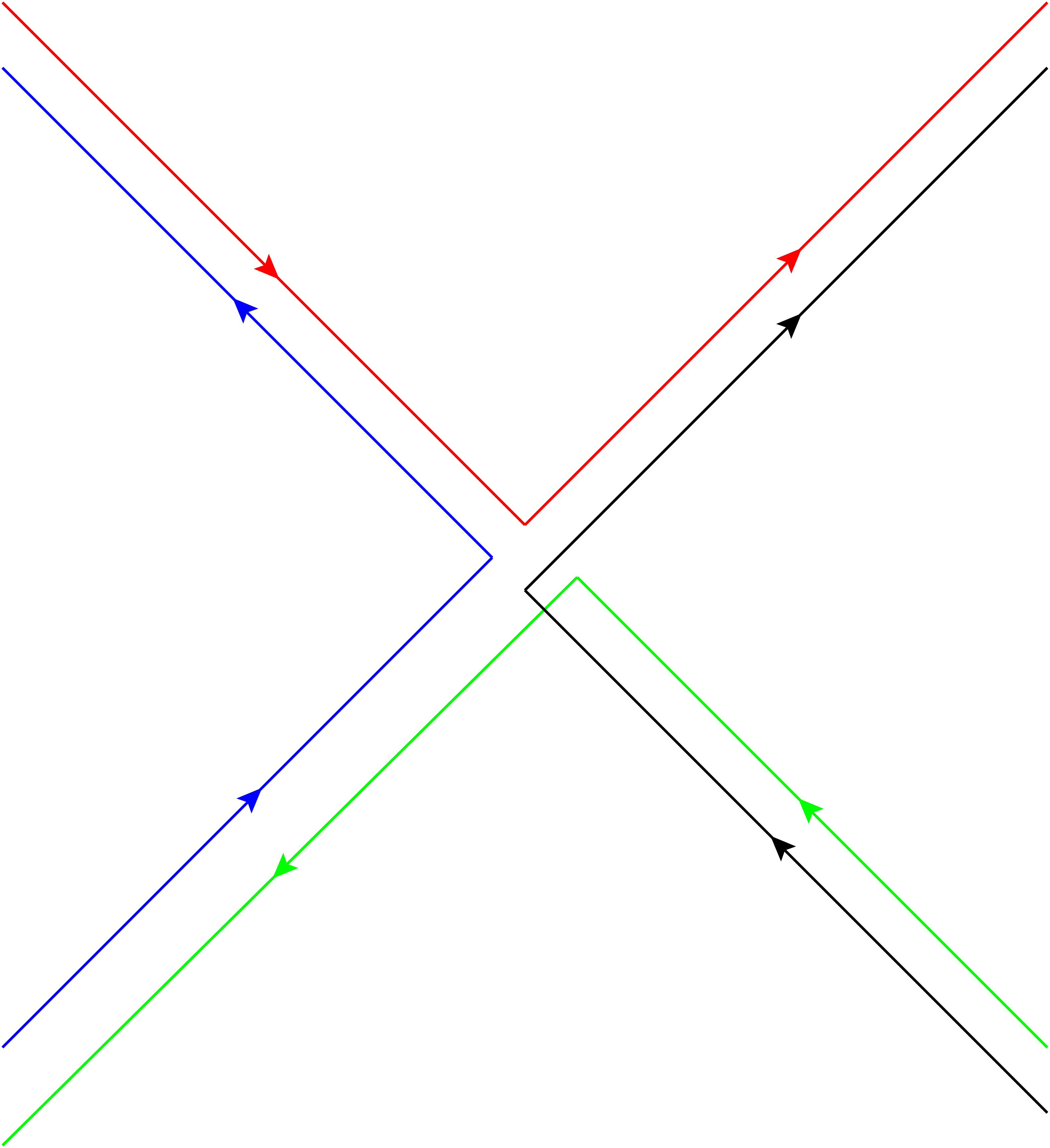}
    }
    \subfigure[]{
      \includegraphics[width=0.27\textwidth]{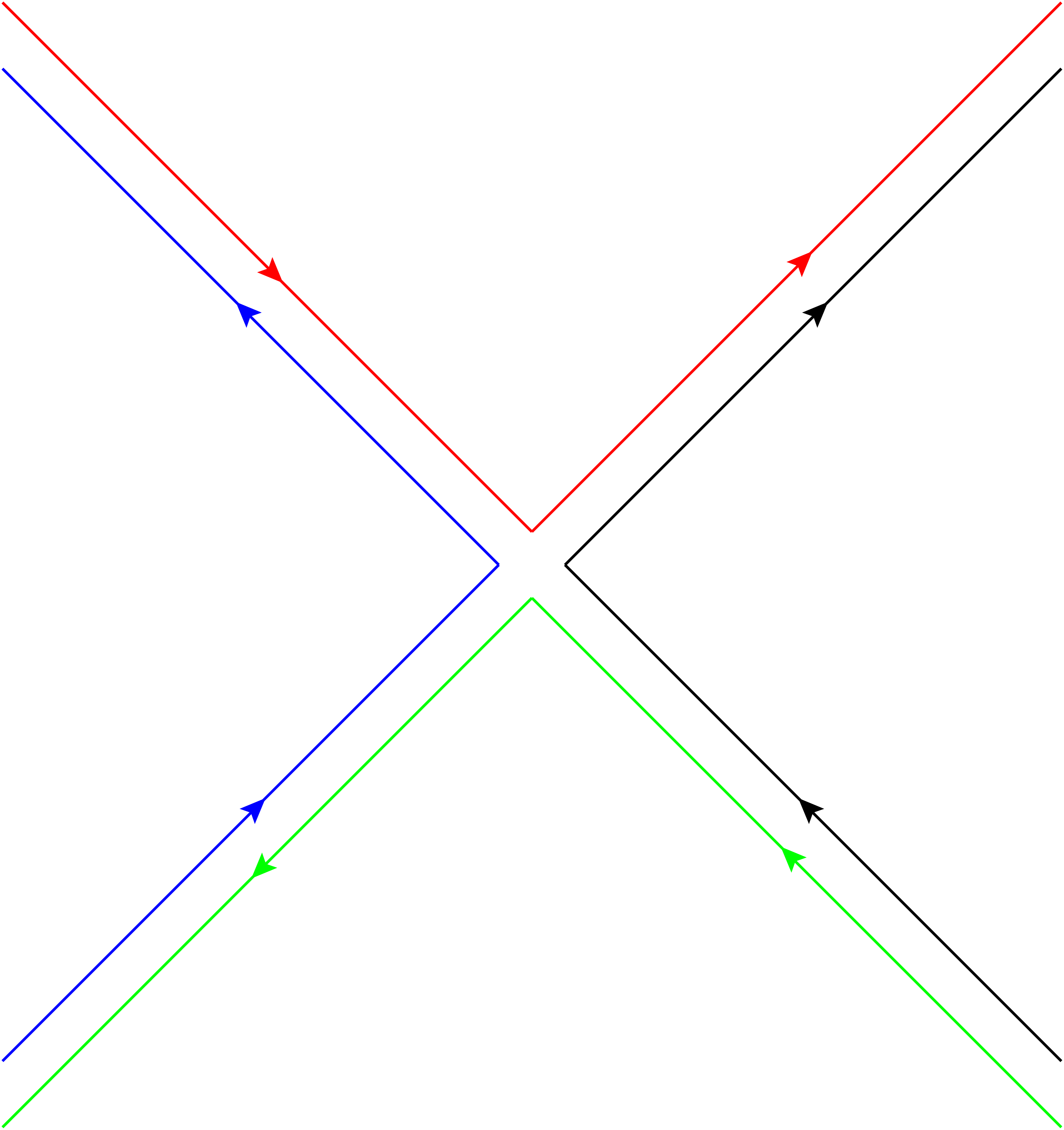}
    }
    \subfigure[]{
      \includegraphics[width=0.27\textwidth]{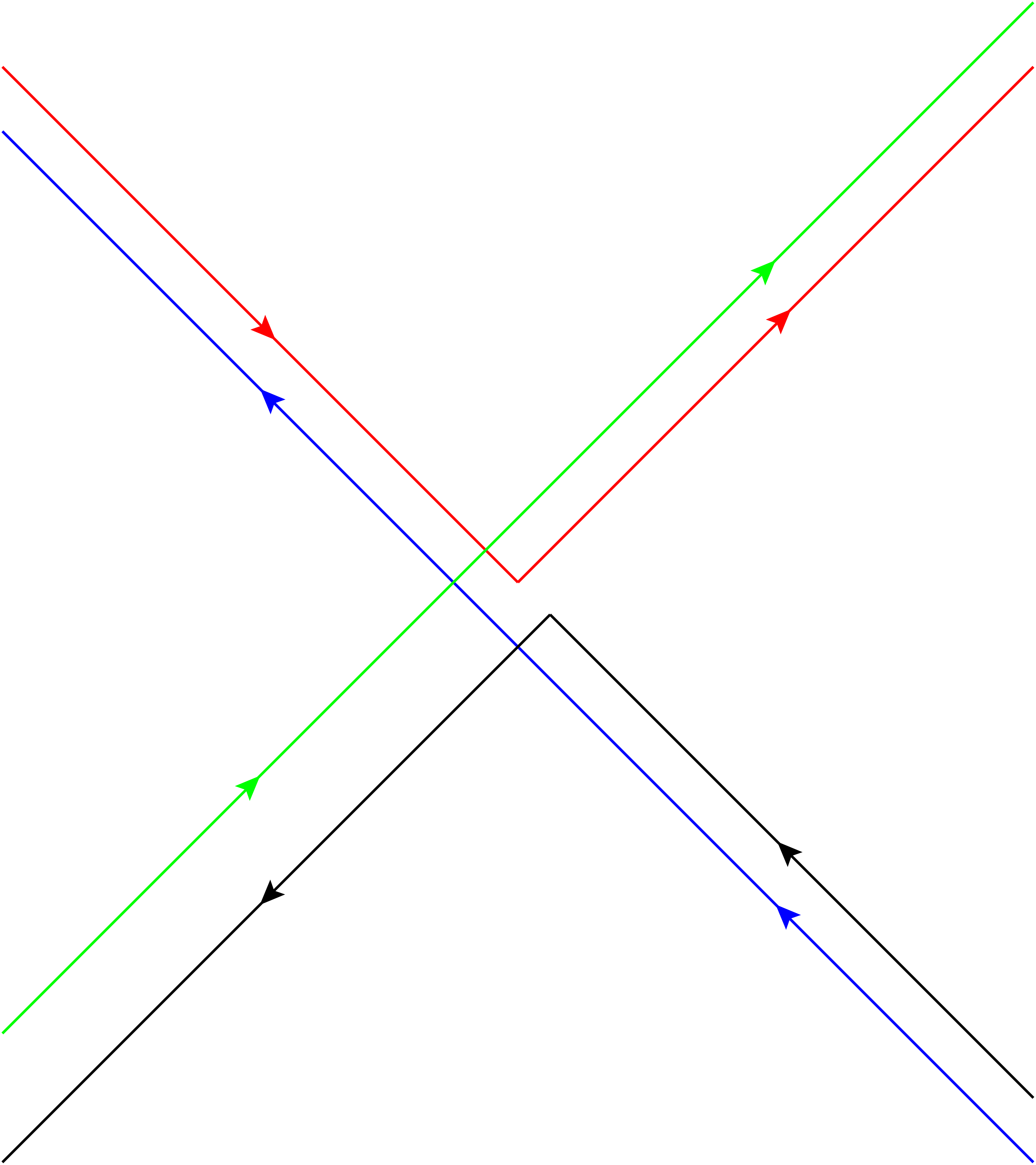}
    }
    \subfigure[]{
      \includegraphics[width=0.27\textwidth]{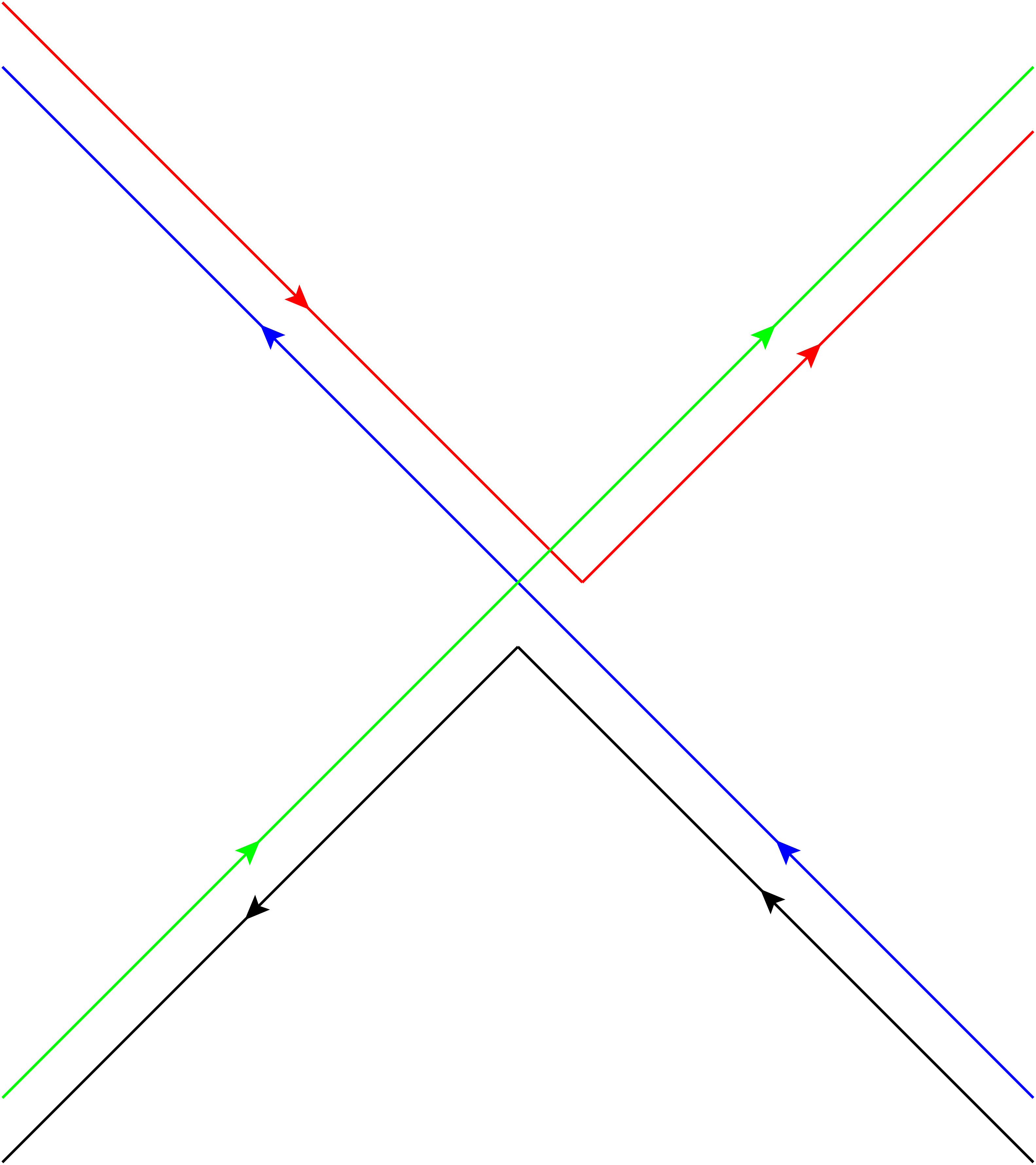}
    }
    \subfigure[]{
      \includegraphics[width=0.27\textwidth]{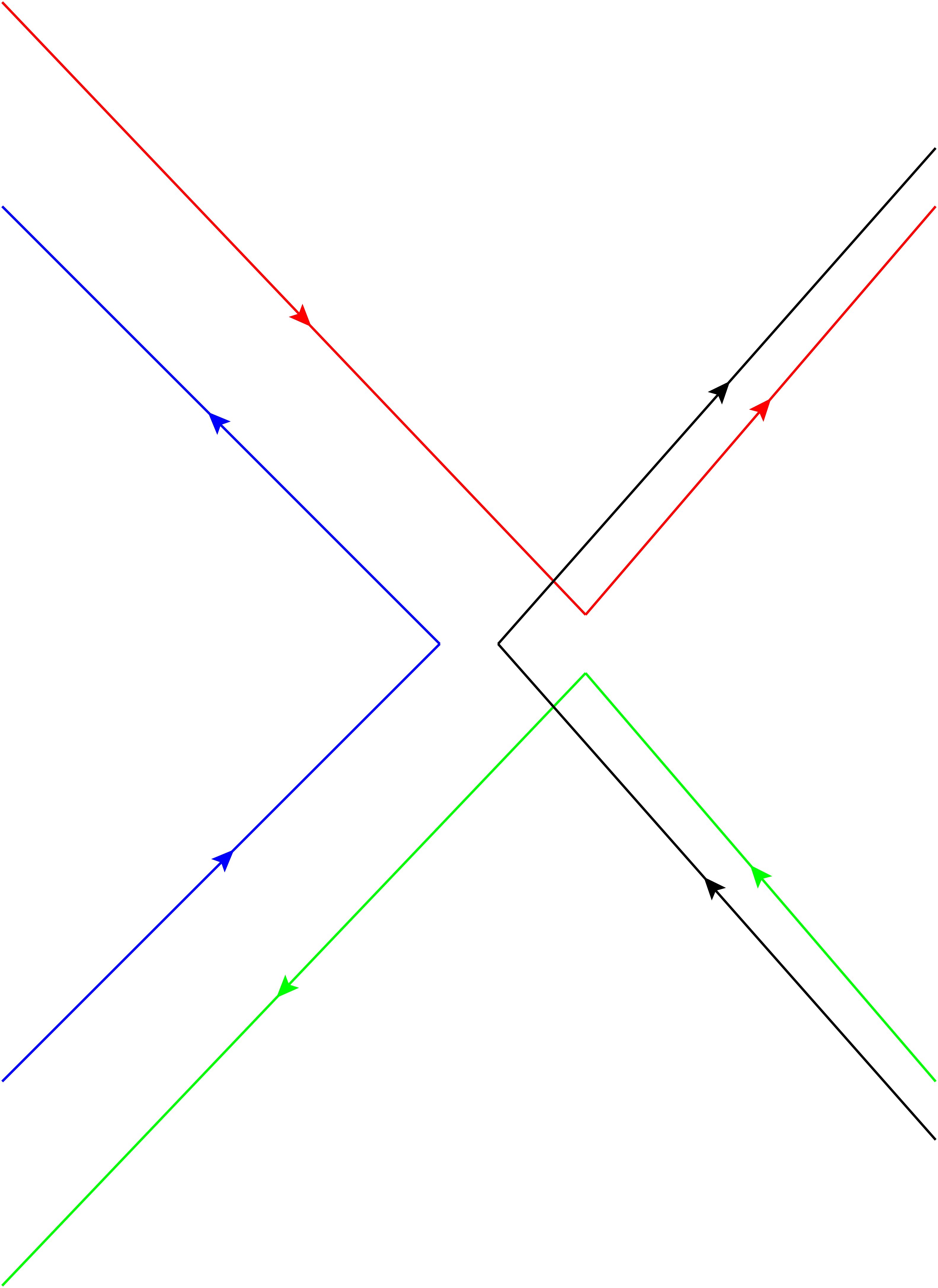}
    }
    \subfigure[]{
      \includegraphics[width=0.27\textwidth]{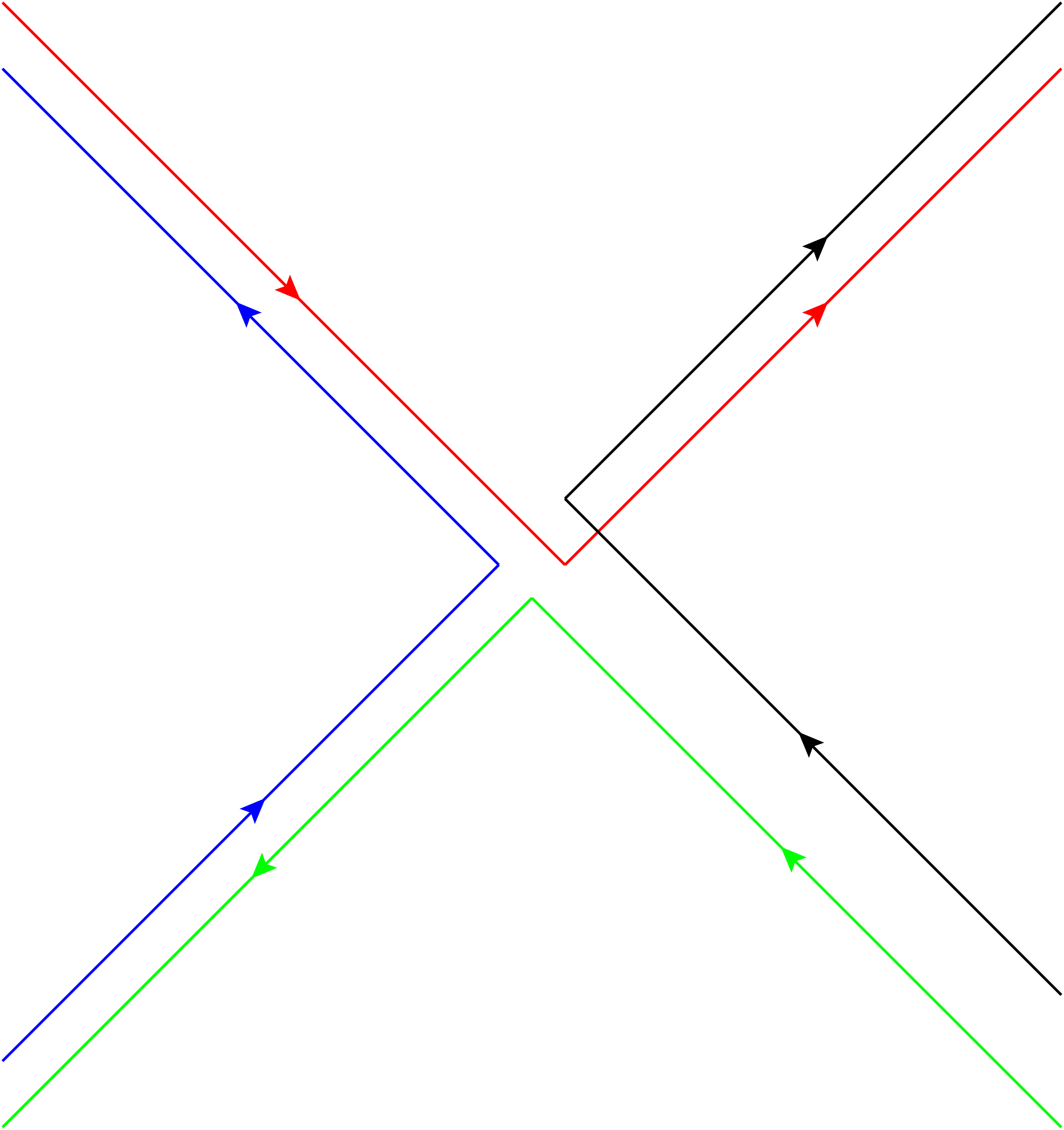}
    }
    \subfigure[]{
      \includegraphics[width=0.27\textwidth]{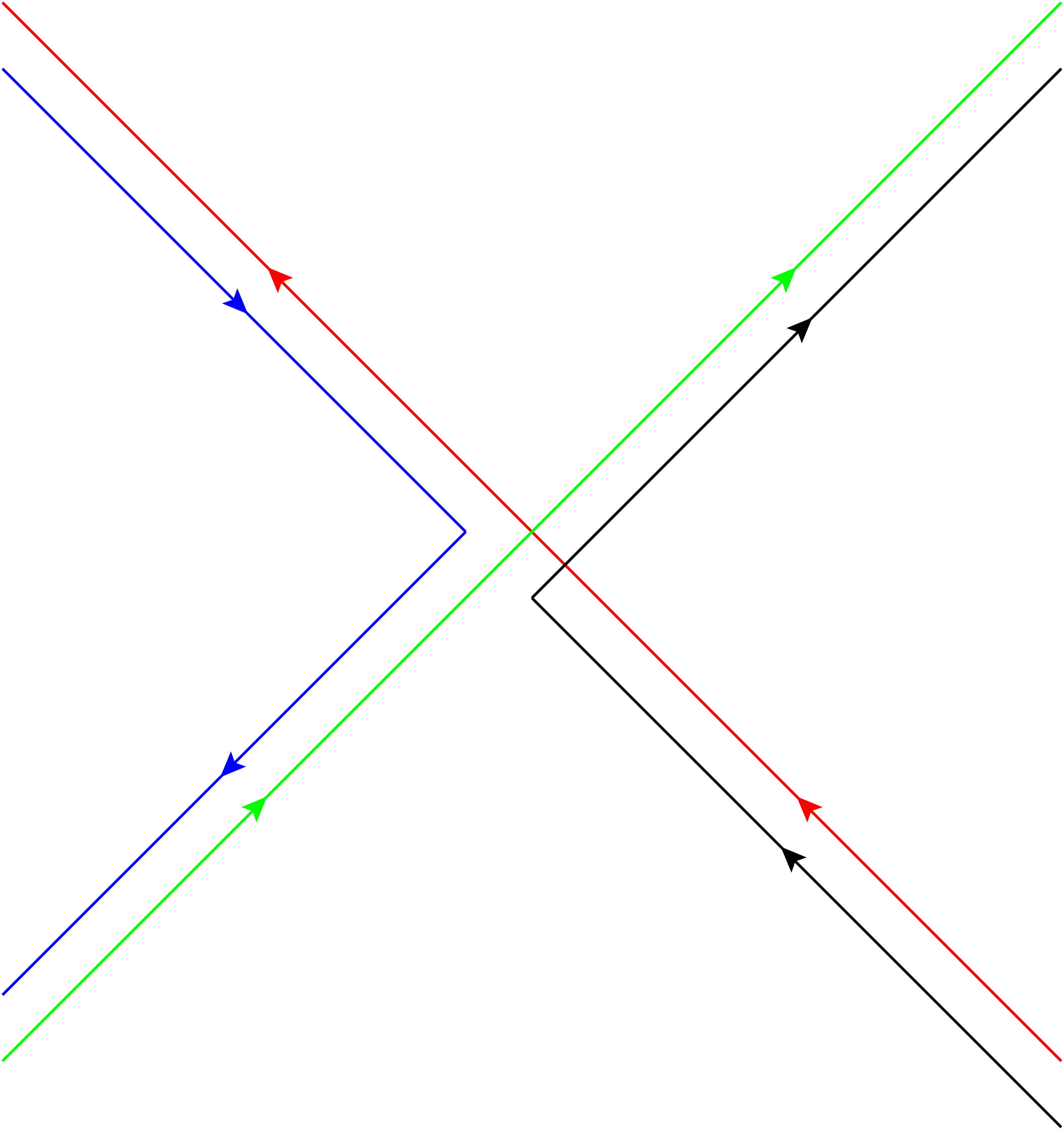}
    }
    \subfigure[]{
      \includegraphics[width=0.27\textwidth]{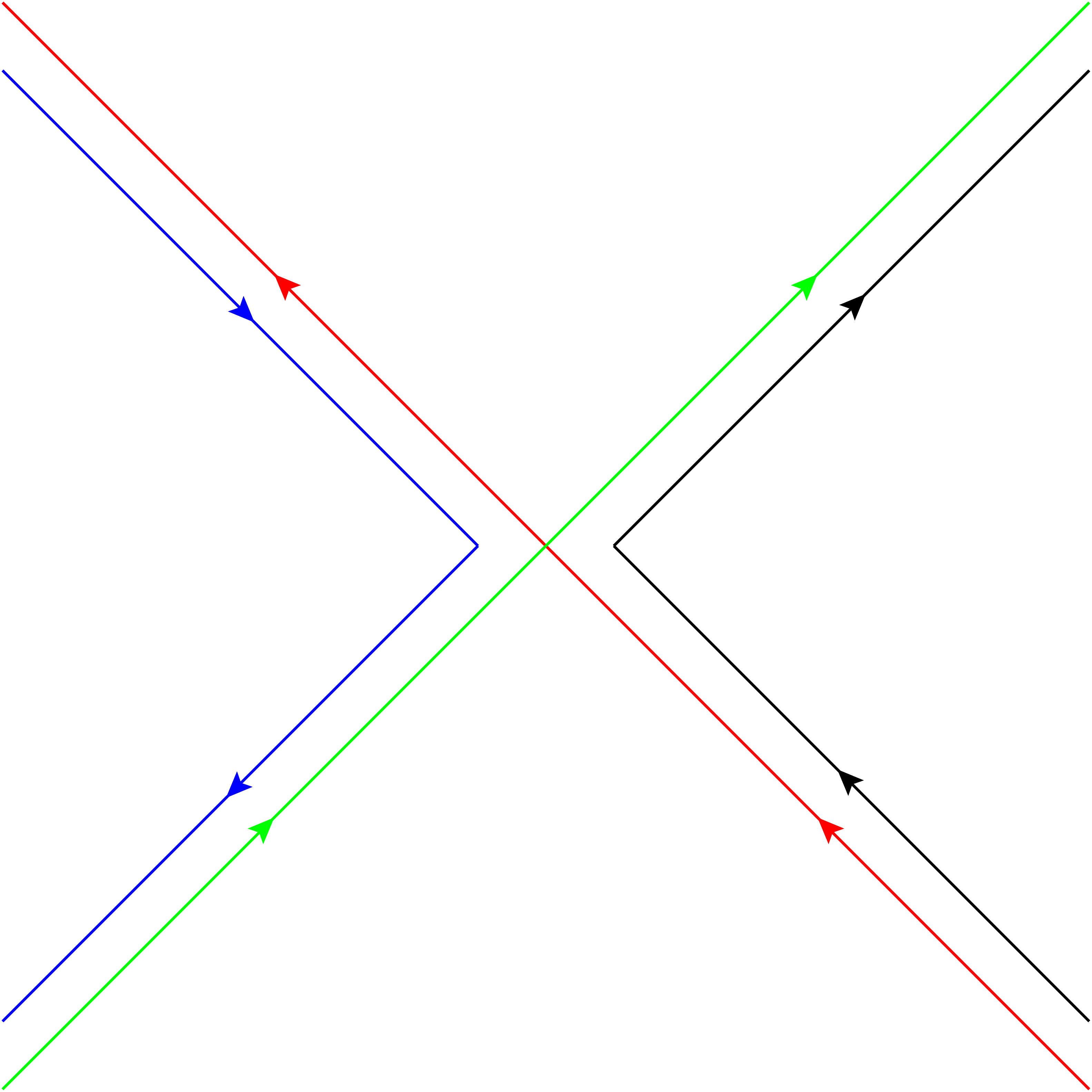}
    }
    \subfigure[]{
      \includegraphics[width=0.27\textwidth]{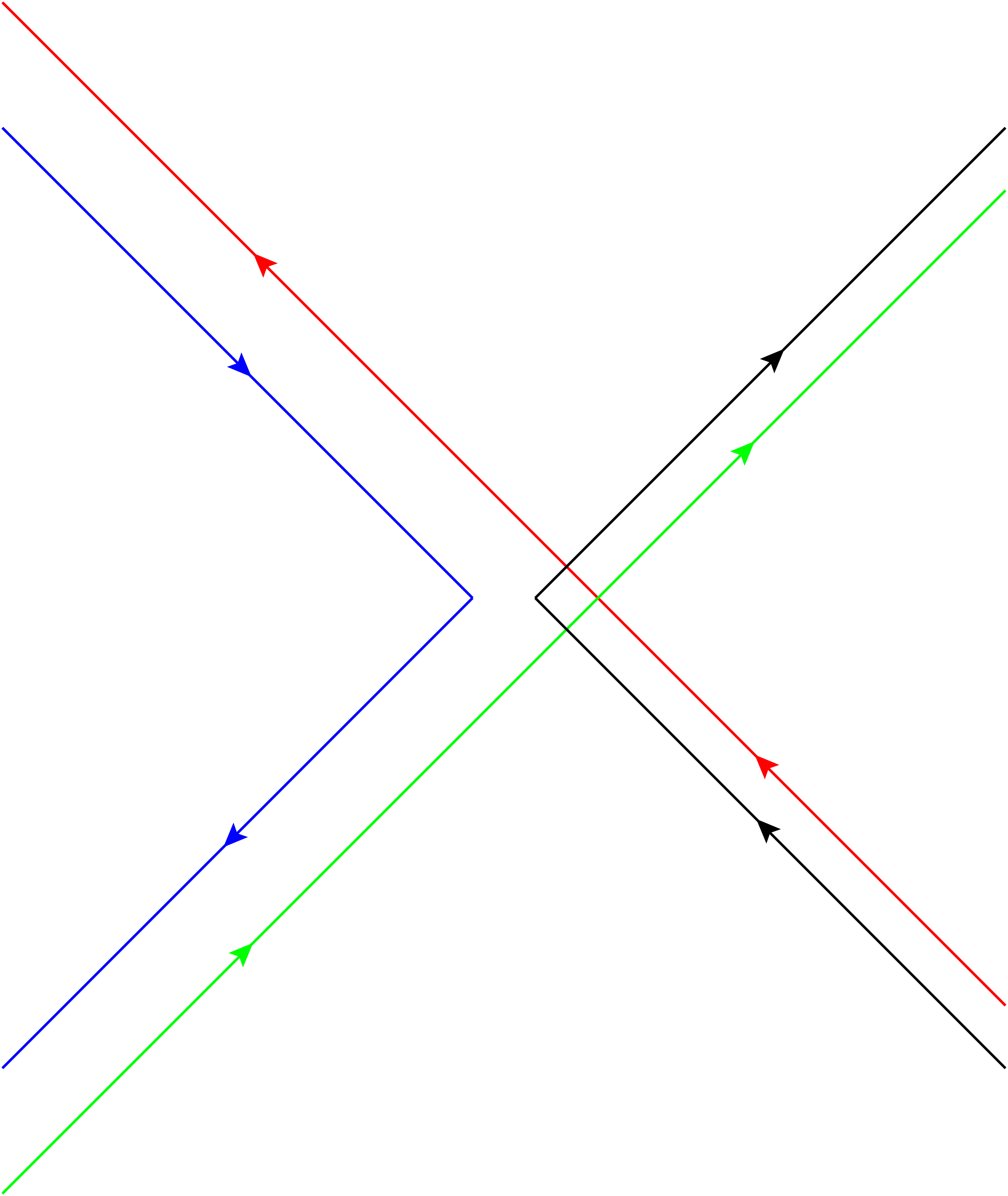}
    }
    \subfigure[]{
      \includegraphics[width=0.27\textwidth]{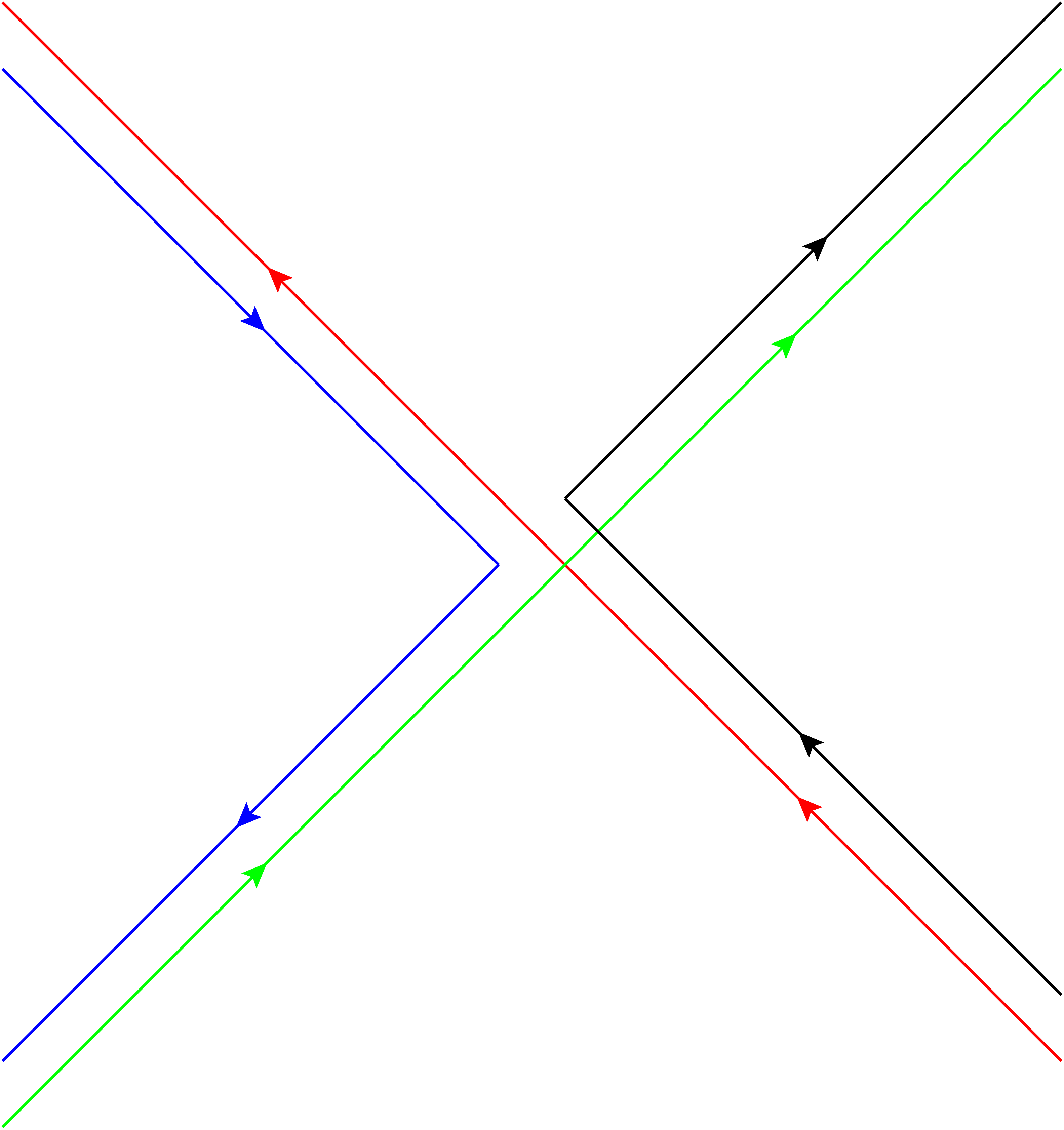}
    }
    \caption{Unique colour flows associated with the 
      pair production of diquarks.}
    \label{fig:pairColourFlows}
  \end{center}
\end{figure}

The pair production process has twelve unique colour flows, as
shown in Figure~\ref{fig:pairColourFlows}. The colours in
Figure~\ref{fig:pairColourFlows} have no physical meaning and are
included as a visual aid.

From these colour flows, Table~\ref{tab:pairColourFlows} is
produced, where $c_1 = \tfrac{(N_C^2-1)^2}{16}$, $c_2 =\tfrac{(N_C^2-1)^2}{16N_C}$,
$c_3 = \tfrac{(N_C^2-1)}{16}$, $c_4 = \tfrac{(N_C^2-1)}{16N_C}$, 
$c_5 = \tfrac{-(N_C^2-1)}{16}$ and $c_6 = \tfrac{-(N_C^2-1)}{16N_C}$.
\\
\begin{table}[h]
  \begin{center}
    \begin{tabular}[c]{c ||c |c |c |c |c |c |c |c |c |c |c |c}
      Diagram & (a) & (b) & (c) & (d) & (e) & (f) & (g) & (h) & (i) & (j) & (k) & (l) \\
      \hline \hline
      (a)$\dagger$ & $c_1$ & $c_2$ & $c_2$ & $0$ & $c_2$ & $c_3$ & $0$ & $c_6$ & $c_6$ & $0$ & $0$ & $c_2$ \\
      \hline
      (b)$\dagger$ & $c_2$ & $c_1$ & $0$ & $c_2$ & $c_3$ & $c_2$ & $c_6$ & $0$ & $0$ & $c_6$ & $c_2$ & $0$ \\
      \hline
      (c)$\dagger$ & $c_2$ & $0$ & $c_1$ & $c_2$ & $0$ & $c_6$ & $c_2$ & $c_3$ & $c_5$ & $c_5$ & $c_5$ & $c_3$ \\
      \hline
      (d)$\dagger$ & $0$ & $c_2$ & $c_2$ & $c_1$ & $c_6$ & $0$ & $c_3$ & $c_4$ & $c_6$ & $c_5$ & $c_3$ & $c_6$ \\
      \hline
      (e)$\dagger$ & $c_2$ & $c_5$ & $0$ & $c_6$ & $c_1$ & $c_1$ & $c_2$ & $0$ & $0$ & $c_2$ & $c_6$ & $0$ \\
      \hline
      (f)$\dagger$ & $c_3$ & $c_2$ & $c_6$ & $0$ & $c_1$ & $c_1$ & $0$ & $c_2$ & $c_2$ & $0$ & $0$ & $c_6$ \\
      \hline
      (g)$\dagger$ & $0$ & $c_6$ & $c_4$ & $c_3$ & $c_2$ & $0$ & $c_1$ & $c_2$ & $c_6$ & $c_3$ & $c_5$ & $c_6$ \\
      \hline
      (h)$\dagger$ & $c_6$ & $0$ & $c_3$ & $c_4$ & $0$ & $c_2$ & $c_2$ & $c_1$ & $c_3$ & $c_6$ & $c_6$ & $c_5$ \\
      \hline
      (i)$\dagger$ & $c_6$ & $0$ & $c_5$ & $c_6$ & $0$ & $c_2$ & $c_6$ & $c_3$ & $c_1$ & $c_2$ & $c_2$ & $c_3$ \\
      \hline
      (j)$\dagger$ & $0$ & $c_6$ & $c_5$ & $c_5$ & $c_2$ & $0$ & $c_3$ & $c_6$ & $c_2$ & $c_1$ & $c_3$ & $c_2$ \\
      \hline
      (k)$\dagger$ & $0$ & $c_2$ & $c_5$ & $c_3$ & $c_6$ & $0$ & $c_5$ & $c_6$ & $c_2$ & $c_3$ & $c_1$ & $c_2$ \\
      \hline 
      (l)$\dagger$ & $c_2$ & $0$ & $c_3$ & $c_6$ & $0$ & $c_6$ & $c_6$ & $c_5$ & $c_3$ & $c_2$ & $c_2$ & $c_1$ \\
    \end{tabular}
    \caption{Associated colour factors for the diagrams shown in Figure~\ref{fig:pairColourFlows}.
      The values of $c_i$ are given in the text.}
    \label{tab:pairColourFlows}
  \end{center}
\end{table}

\subsection{Shower}
\label{app:shower}

\begin{figure}[h]
  \begin{center}
    \subfigure[]{
      \includegraphics[width=0.3\textwidth]{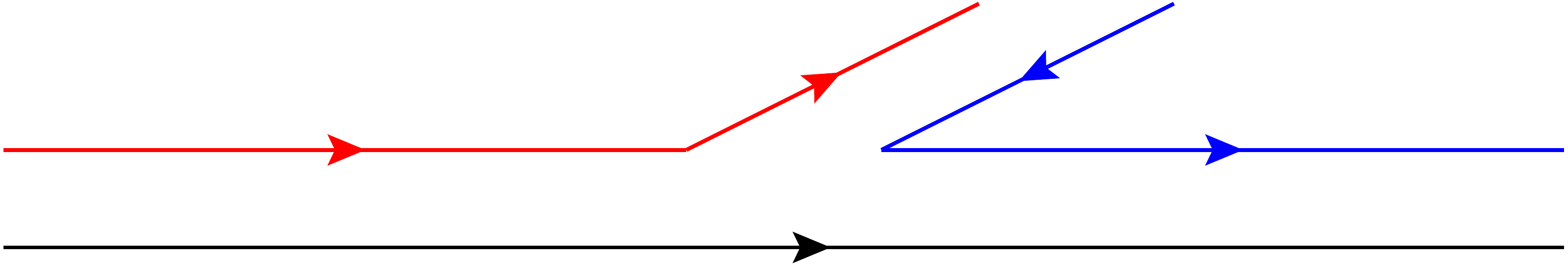}
      \label{subfig:showerA}
    }
    \subfigure[]{
      \includegraphics[width=0.3\textwidth]{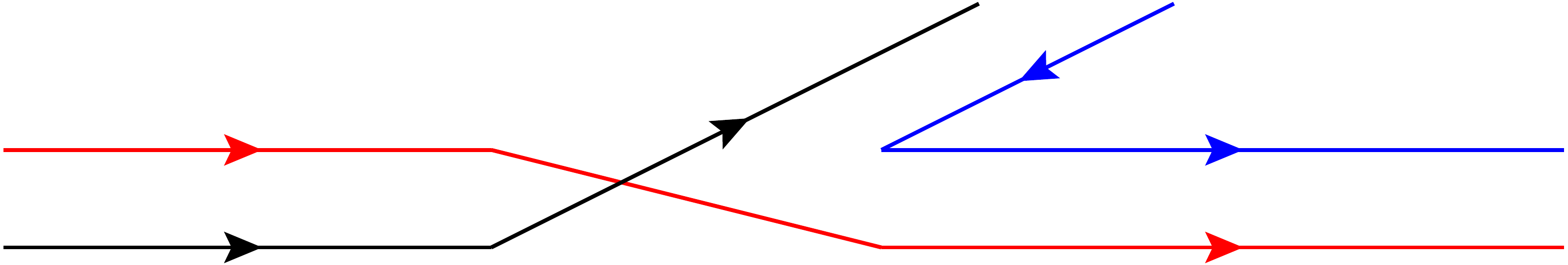}
      \label{subfig:showerC}
    }
    \\
    \subfigure[]{
      \includegraphics[width=0.3\textwidth]{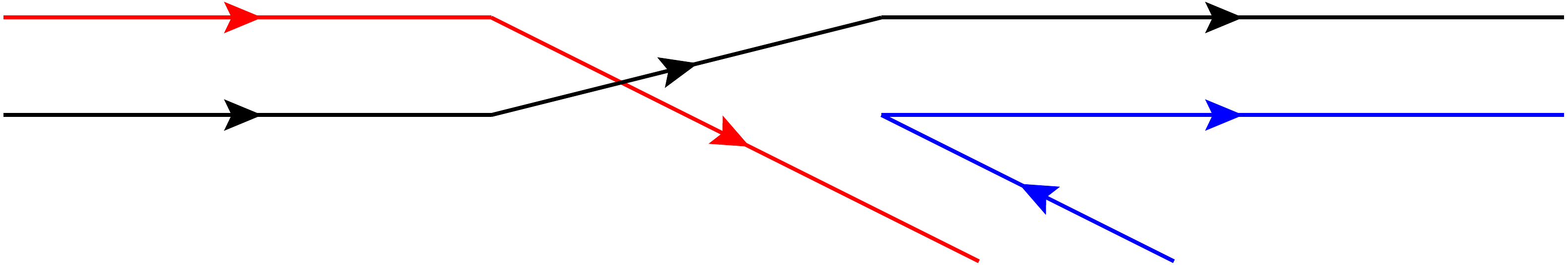}
      \label{subfig:showerB}
    }
    \subfigure[]{
      \includegraphics[width=0.3\textwidth]{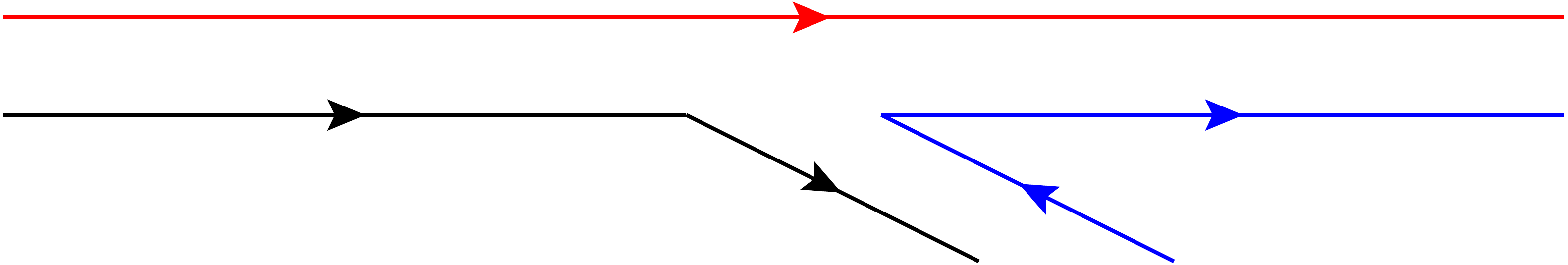}
      \label{subfig:showerD}
    }
    \caption{Colour flows for a diquark emitting a gluon during the
      shower.}
    \label{fig:showerColour}
  \end{center}
\end{figure}

The splitting functions were decomposed in the same way as for the
pair production. There are four unique
colour flows associated with a diquark emitting a gluon, as shown in
Figure~\ref{fig:showerColour}, again where the colours are included as
a visual aid.

Only the colour prefactor of the existing splitting functions is
changed.  The colour prefactor is given by $\tfrac{10}{3}$, i.e. the
diquarks radiate $2\frac12$ times more than a particle in the octet representation.

During the shower, it is assumed that gluons did not branch to form
diquarks owing to the large diquark mass.

\bibliography{Herwig++}
\end{document}